\documentclass[14pt, preprint2]{aastex6}
\usepackage{graphicx, subfigure, algorithm, algorithmic, amsmath, nicefrac, hyperref, enumitem}
\hypersetup{allcolors = [rgb]{0., 0., 0.5}} 

\usepackage{longtable}
\newcommand{\project}[1]{\textsl{#1}}

\newcommand{\apogee}{\project{\textsc{apogee}}}
\newcommand{\apokasc}{\project{\textsc{apokasc}}}
\newcommand{\aspcap}{\project{\textsc{aspcap}}}

\newcommand{\sloanv}{\project{Sloan V}}
\newcommand{\mwm}{\project{Milky Way Mapper}}

\newcommand{\documentname}

\usepackage{booktabs}

\newcommand{\teff}{\mbox{$T_{\rm eff}$}}

\newcommand{\logg}{\mbox{$\log g$}}

\newcommand{\rgal}{\mbox{$R_{\text{GAL}}$}}

\newcommand{\xone}{\ensuremath{x_n}}
\newcommand{\xtwo}{\ensuremath{x_{n'}}}

\begin{document}

\title{The Age Distribution of Stars in the Milky Way Bulge}

\author{Tawny Sit\altaffilmark{1}, M.K.~Ness\altaffilmark{2,3}}  

\altaffiltext{1}{California Institute of Technology, Pasadena, CA 91125, USA} \
\altaffiltext{2}{Department of Astronomy, Columbia University, Pupin Physics Laboratories, New York, NY 10027, USA}\
\altaffiltext{3}{Center for Computational Astrophysics, Flatiron Institute, 162 Fifth Avenue, New York, NY 10010, USA}

\email{tsit@alumni.caltech.edu}

\begin{abstract} 
The age and chemical characteristics of the Galactic bulge link to the formation and evolutionary history of the Galaxy.  Data-driven methods and large surveys enable stellar ages and precision chemical abundances to be determined for vast regions of the Milky Way, including the bulge. Here, we use the data-driven approach of \textit{The Cannon}, to infer the ages and abundances for 125,367 stars in the Milky Way, using spectra from Apache Point
Observatory Galaxy Evolution Experiment (\apogee) DR14. We examine the ages and metallicities of 1654 bulge stars within $\rgal<3.5$ kpc. We focus on fields with $b<12^\circ$, and out to longitudes of $l<15^\circ$. We see that stars in the bulge are about twice as old ($\tau=8$ Gyr), on average, compared to those in the solar neighborhood ($\tau=4$ Gyr), with a larger dispersion in [Fe/H] ($\approx0.38$ compared to 0.23 dex). This age gradient comes primarily from the low-$\alpha$ stars. Looking along the Galactic plane, the very central field in the bulge shows by far the largest dispersion in [Fe/H] ($\sigma_{[Fe/H]}\approx0.4$ dex) and line-of-sight velocity ($\sigma_{vr}\approx90$ km/s), and simultaneously the smallest dispersion in age. Moving out in longitude, the stars become kinematically colder and less dispersed in [Fe/H], but show a much broader range of ages. We see a signature of the X-shape within the bulge at a latitude of $b=8^\circ$, but not at $b=12^\circ$. Future \apogee\ and other survey data, with larger sampling, affords the opportunity to extend our approach and study in more detail, to place stronger constraints on models of the Milky Way.

\end{abstract}

\section{Introduction}

The Milky Way is a barred spiral galaxy with a boxy/peanut bulge morphology and an X-shaped structure \citep[see][and references therein]{Go2016}. Photometric measurements of red clump stars, which exhibit two density peaks along our line of sight \citep[e.g.][]{2McW2010, Nataf2010}, and \textit{N}-body simulations which show the formation of boxy/peanut and X-shaped bulges via dynamical instabilities in the disk \citep[e.g.][]{Ness2012}, both provide evidence of the boxy, X-shaped bulge. This shape can be seen in the Milky Way from the stellar brightness alone \citep{Ness2016b} and indicates that the bulge was, at least in large part, formed from the disk, via dynamical instabilities \citep[e.g.][]{Athanassoula2005, Debattista2006, MartinezValpuesta2006}. Metallicity maps of the bulge have suggested that the observed properties of the boxy/peanut bulge are a consequence of the initial properties of a two-component (thin and thick) disk \citep[e.g.][]{Gonzalez2013, diM2015, Fragkoudi2018}. More generally, the disk stellar populations can separate based on their kinematics \citep[see][]{Debattista2017} and today the bulge shows correlations between chemistry and kinematics \citep[see][and references therein]{Babusiaux2016}. With new large surveys across the Galaxy, we have the opportunity to extensively map the bulge and surrounding disk, with respect to stellar age in particular \citep[e.g.]{Bovy2019}, which has been shown to vary in the bulge, with younger stars more strongly exhibiting the X-shape than older stars \citep[e.g.][]{Debattista2017,Grady2020}.

\begin{figure*}[!ht]
    \centering
    \includegraphics[scale=0.42, keepaspectratio]{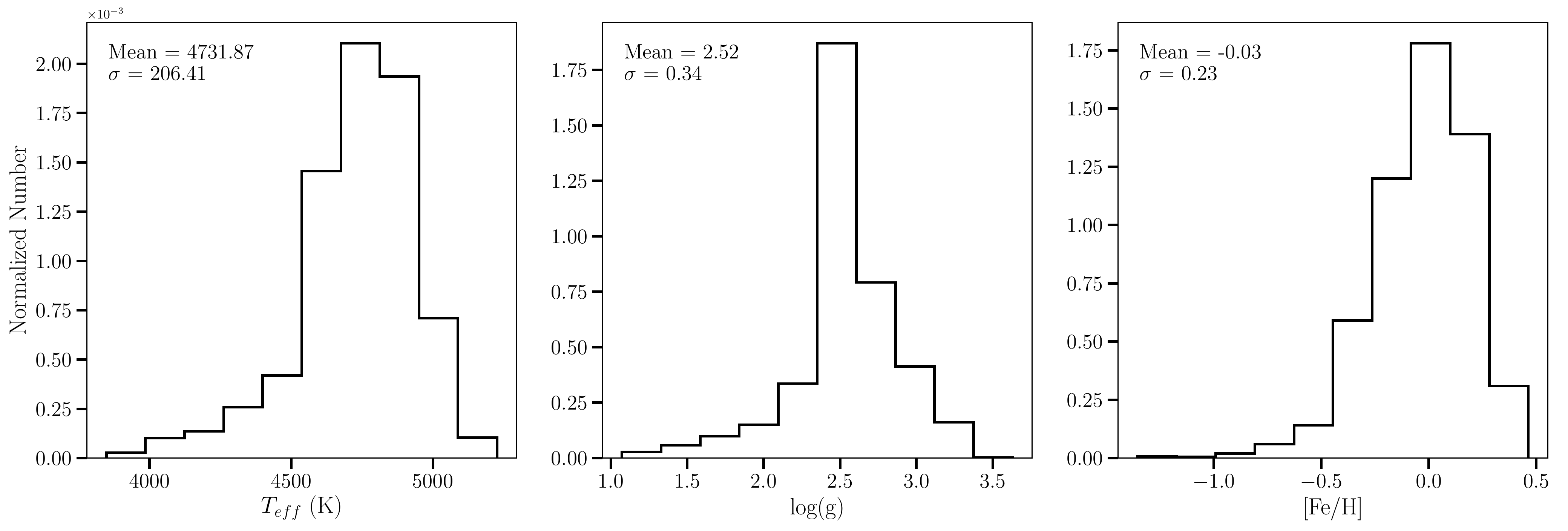}
    \caption{\teff, $\log{g}$, [Fe/H] distributions of the the full training set of 5991 \apokasc\ stars.}
    \label{fig: label space histograms}
\end{figure*}

The Apache Point Observatory Galaxy Evolution Experiment (\apogee) survey has observed $\sim$470,000 stars in the Milky Way \citep{Majewski2017}, as released in DR16 \citep{Ahumada2019}. In this work, we use the $\sim$280,000 stars released in DR14 \citep{Blanton2017}. \apogee\ has obtained high-resolution $R=22,500$ spectra of these stars \citep{Zasowski2013,Zasowski2017} in the near-infrared ($H$-band). Observing in the $H$-band allows for better observations of stars obscured by dust than observing in the optical band, making it an ideal instrument for observing the far Galactic disk and bulge. Relevant to our work, \apogee\ has observed a few special fields of stars in the outer arms of the X-shaped bulge, as described in \citet{Zasowski2017}. 
\par
\textit{The Cannon} is a data-driven approach for measuring stellar parameters and abundances, collectively called labels, from stellar spectra \citep{Ness2015}. Such data-driven methods are, in general, able to learn a relationship between line strength and labels, without necessarily knowing the underlying physics, possibly circumventing any poorly defined or approximated line strength-abundance conversion that the traditional methods rely on. Data-driven models can also find labels for a large dataset with little computational effort. \textit{The Cannon}, in particular, can also derive precision stellar labels at lower signal-to-noise than prior approaches. \textit{The Cannon} uses a set of reference objects (the ``training set"), of stars with spectra that have well-defined, high-fidelity labels, that span the label space of the ``test set", to create a model. This model captures the relationship between the flux at every pixel and the label values. Then, \textit{The Cannon} uses the model to fit spectra from stars with unknown labels (the ``test set"). The first iteration of \textit{The Cannon} has been used to fit 55,000 stars from \apogee\ DR10 with a training set of only 542 reference objects \citep{Ness2015}. Later, 450,000 Large Sky Area Multi-Object Fiber Spectroscopic Telescope (LAMOST) giants were labeled using a training set of 9952 reference objects, taken from the stars in common between the \apogee\ DR12 and LAMOST surveys \citep{Ho2017a,Ho2017b}. A more sophisticated, regularized version of \textit{The Cannon} \citep{Casey2016} has been applied to label the Radial Velocity Experiment survey with precision individual abundance measurements \citep{Casey2016b}.

\section{Data} 

\begin{figure}[!t]
    \centering
    \includegraphics[scale=0.52, keepaspectratio]{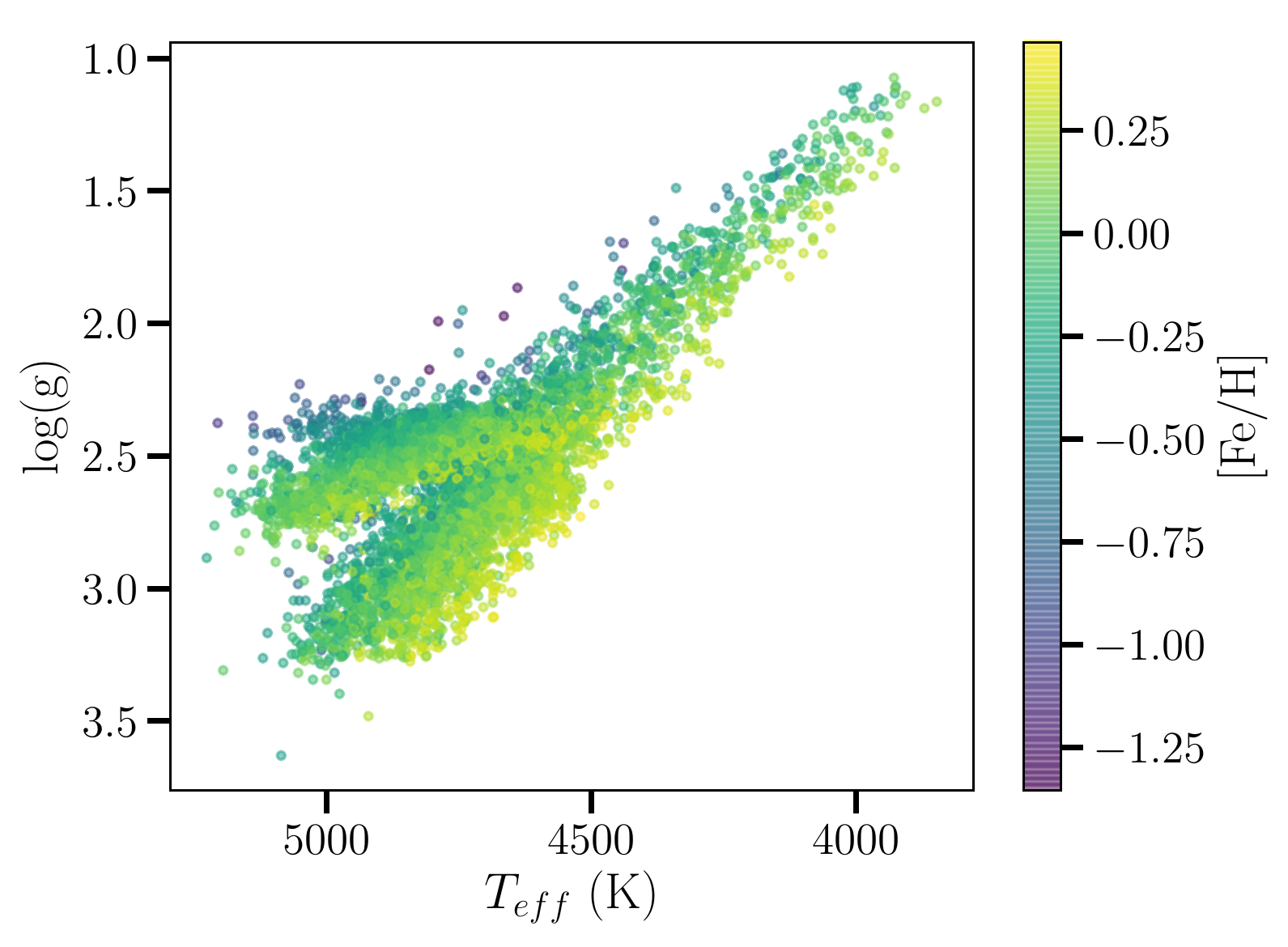}
    \caption{\teff vs. $\log{g}$, colored by [Fe/H] of the training set, which contains a total of 5991 stars.}
    \label{fig: teff-logg scatter plot}
\end{figure}

For the training set, we started by using 6753 stars common to \apogee\ DR14 \citep{Holtzman2018} and Kepler. The labels of the stars are given by spectroscopic parameters from the \apogee\ pipeline, \aspcap\ \citep{GP2016}, and ages from Kepler astroseismic data, in the \apokasc\ catalog \citep{Pins2018}. We clean the full set of common stars by making a series of quality cuts. All stars with an anomalous label in metallicity, or C, N, O, Mg, Al, Si, S, K, Ca, Ti, V, Mn, Ni, P, Cr, or Co abundance were removed. An anomalous label was defined as [Fe/H] or [X/Fe] $>$ 1.5. One star with a very high age error was also removed. Finally, a signal-to-noise cut was made, leaving only stars that had signal-to-noise ratio $>$ 100. After these quality cuts, 5991 stars remained for the final test set. The test set, whose label space is plotted in Figures \ref{fig: label space histograms} and \ref{fig: teff-logg scatter plot}, spans 3848 K $<\teff<$ 5223 K, 1.07 $<\log{g}<$ 3.63, and $-1.35<$ [Fe/H] $<$ 0.46.

\begin{figure}[!th]
    \centering
    \includegraphics[scale=0.54, keepaspectratio]{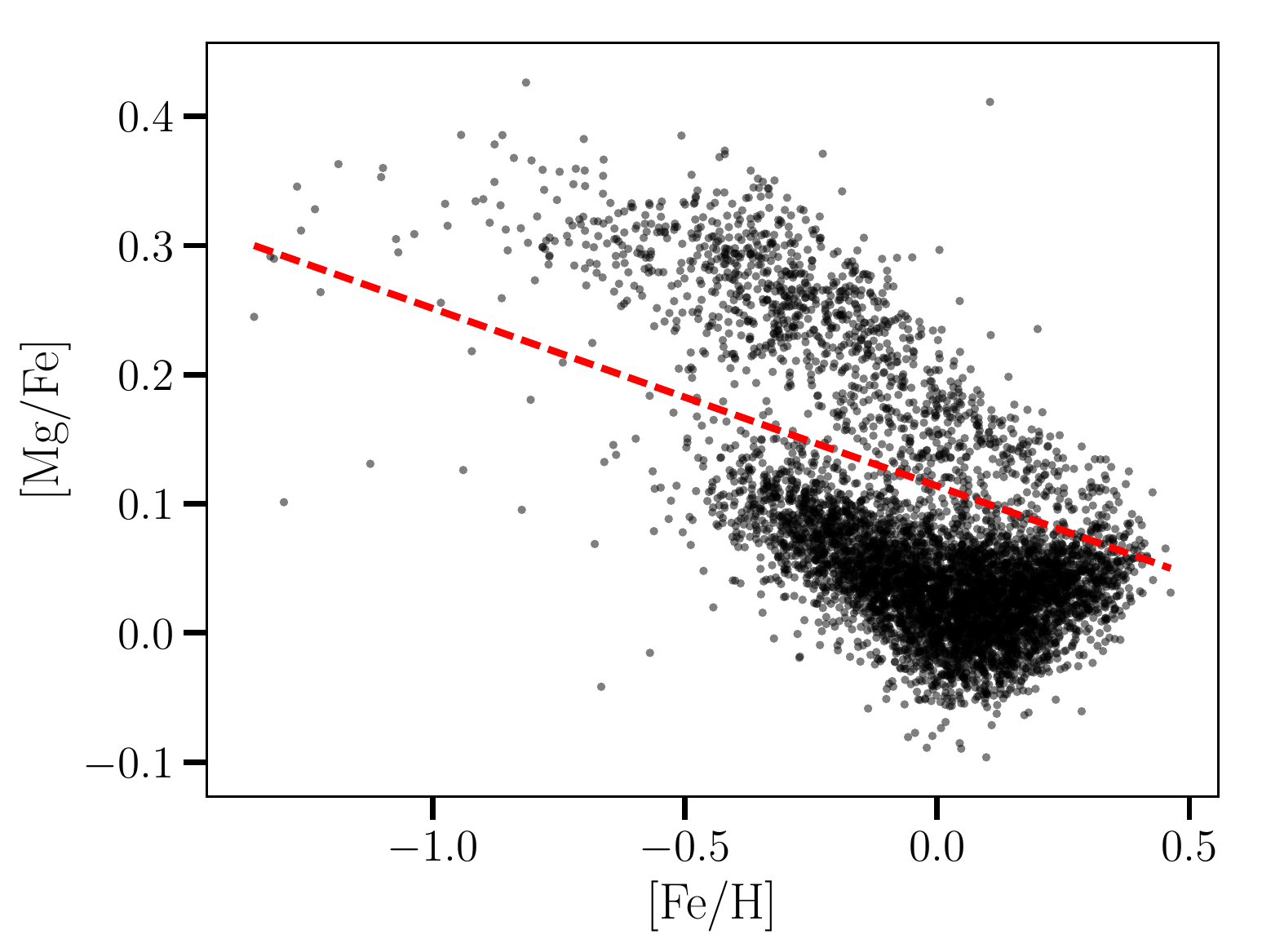}
    \caption{[Fe/H] vs. [Mg/Fe], where [Mg/Fe] is used as a proxy for overall [$\alpha$/Fe], of the training set with total 5991 stars. The cutoff line dividing high-$\alpha$ and low-$\alpha$ stars is shown, dividing the set into 1255 high-$\alpha$ and 4736 low-$\alpha$ training set stars.}
    \label{fig: alpha cutoff}
\end{figure}

\begin{table*}[!ht]
\centering
\caption{High-$\alpha$ Cross-validation Comparison} 
\label{high-alpha cross-validation table}
\begin{tabular}{|l|l|l|l|l|}
\hline
            & Unnormalized Scatter & Normalized Scatter & Unnormalized Bias & Normalized Bias \\ \hline
\teff       & 30.40             & 31.22           & 6.58           & 5.72          \\ \hline
$\log{g}$   & 0.07              & 0.07            & 0.00           & 0.01          \\ \hline
{[}Fe/H{]}  & 0.03              & 0.03            & 0.00            & 0.00         \\ \hline
{[}C/Fe{]}  & 0.05              & 0.05            & 0.00            & 0.00       \\ \hline
{[}N/Fe{]}  & 0.06              & 0.06            & 0.00           & 0.00          \\ \hline
{[}O/Fe{]}  & 0.04              & 0.04            & 0.00           & 0.00         \\ \hline
{[}Mg/Fe{]} & 0.03              & 0.03            & 0.00          & 0.00        \\ \hline
{[}Al/Fe{]} & 0.07              & 0.07            & 0.00           & 0.00         \\ \hline
{[}Si/Fe{]} & 0.03              & 0.03            & 0.00           & 0.00         \\ \hline
{[}K/Fe{]}  & 0.06              & 0.06            & 0.00          & 0.00        \\ \hline
{[}Ca/Fe{]} & 0.03              & 0.03            & 0.00           & 0.00        \\ \hline
{[}Mn/Fe{]} & 0.04              & 0.04             & 0.00           & 0.00         \\ \hline
log(age) & 0.20              & 0.21            & 0.00          & 0.00         \\ \hline
\end{tabular}
\end{table*}

\begin{table*}[!ht]
\centering
\caption{Low-$\alpha$ Cross-validation Comparison}
\label{low-alpha cross-validation table}
\begin{tabular}{|l|l|l|l|l|}
\hline
            & Unnormalized Scatter & Normalized Scatter & Unnormalized Bias & Normalized Bias \\ \hline
\teff       & 24.64              & 31.73              & 2.85               & 0.79               \\ \hline
$\log{g}$   & 0.07               & 0.08               & 0.00               & 0.00               \\ \hline
{[}Fe/H{]}  & 0.02               & 0.02               & 0.00               & 0.00               \\ \hline
{[}C/Fe{]}  & 0.04               & 0.05               & 0.00               & 0.00              \\ \hline
{[}N/Fe{]}  & 0.05               & 0.06               & 0.00               & 0.00               \\ \hline
{[}O/Fe{]}  & 0.04               & 0.05               & 0.00               & 0.00              \\ \hline
{[}Mg/Fe{]} & 0.02               & 0.03               & 0.00               & 0.00              \\ \hline
{[}Al/Fe{]} & 0.07               & 0.08               & 0.00               & 0.01              \\ \hline
{[}Si/Fe{]} & 0.02               & 0.03               & 0.00               & 0.00               \\ \hline
{[}Ca/Fe{]} & 0.03               & 0.03               & 0.00               & 0.00              \\ \hline
{[}Mn/Fe{]} & 0.03               & 0.04               & 0.00               & 0.00              \\ \hline
log(age) & 0.20                  & 0.22               & -0.01              & -0.02              \\ \hline
\end{tabular}
\end{table*}

\textit{The Cannon}\footnote{For our derivation of stellar labels for our spectra, we use \textit{The Cannon} version available at \url{https://github.com/annayqho/TheCannon}, \citep[see][]{Ho2017a}.} requires that its input spectra be normalized in a way that is independent of signal-to-noise. While the \apogee\ spectra we are using are already normalized, an additional signal-to-noise independent normalization was done on all of the spectra used. In this work, the high- and low-$\alpha$ stars were divided into separate training sets. For both training sets, the spectra were first divided into three wavelength ranges: 15164-15794, 15871-16417, and 16485-16936 \AA. This accounted for different parts of the spectra behaving differently - these ranges roughly correspond to the chips of the spectrograph. Continuum pixels were then identified, as those which had median flux over all spectra close to 1 and little dependence on \teff, $\log{g}$, or [Fe/H]. The flux cut was chosen to select only pixels with a median flux between 0.985 and 1.15. The lack of label dependence was determined by selecting the 30\% of pixels with linear coefficients in each stellar parameter label (\teff, \logg, [Fe/H]) closest to 0 in the model generated by the \aspcap\ spectra prior to our additional continuum normalization step. Pixels that satisfied all of these requirements were considered to be continuum. For the high-$\alpha$ training set, this resulted in 352 pixels, about 4.1\% of all pixels, and for the low-$\alpha$ set, this resulted in 373 pixels, about 4.3\% of all pixels. The continuum pixels were were fit by a third-order Chebyshev function, then all training spectra were divided by this function, to obtain the normalized spectra.
\par
A 10-fold cross validation of the model (see Section \ref{sec: methods}) was created from the newly normalized training spectra along with an unnormalized control case to compare the performance. This resulted in the unnormalized case performing better by a few percent (see Tables \ref{high-alpha cross-validation table} and \ref{low-alpha cross-validation table}). However, we still choose to adopt the normalization because it should perform better (consistently) at low signal to noise.

\begin{figure*}[!hp]
    \centering
    \includegraphics[width=0.9\textwidth, keepaspectratio]{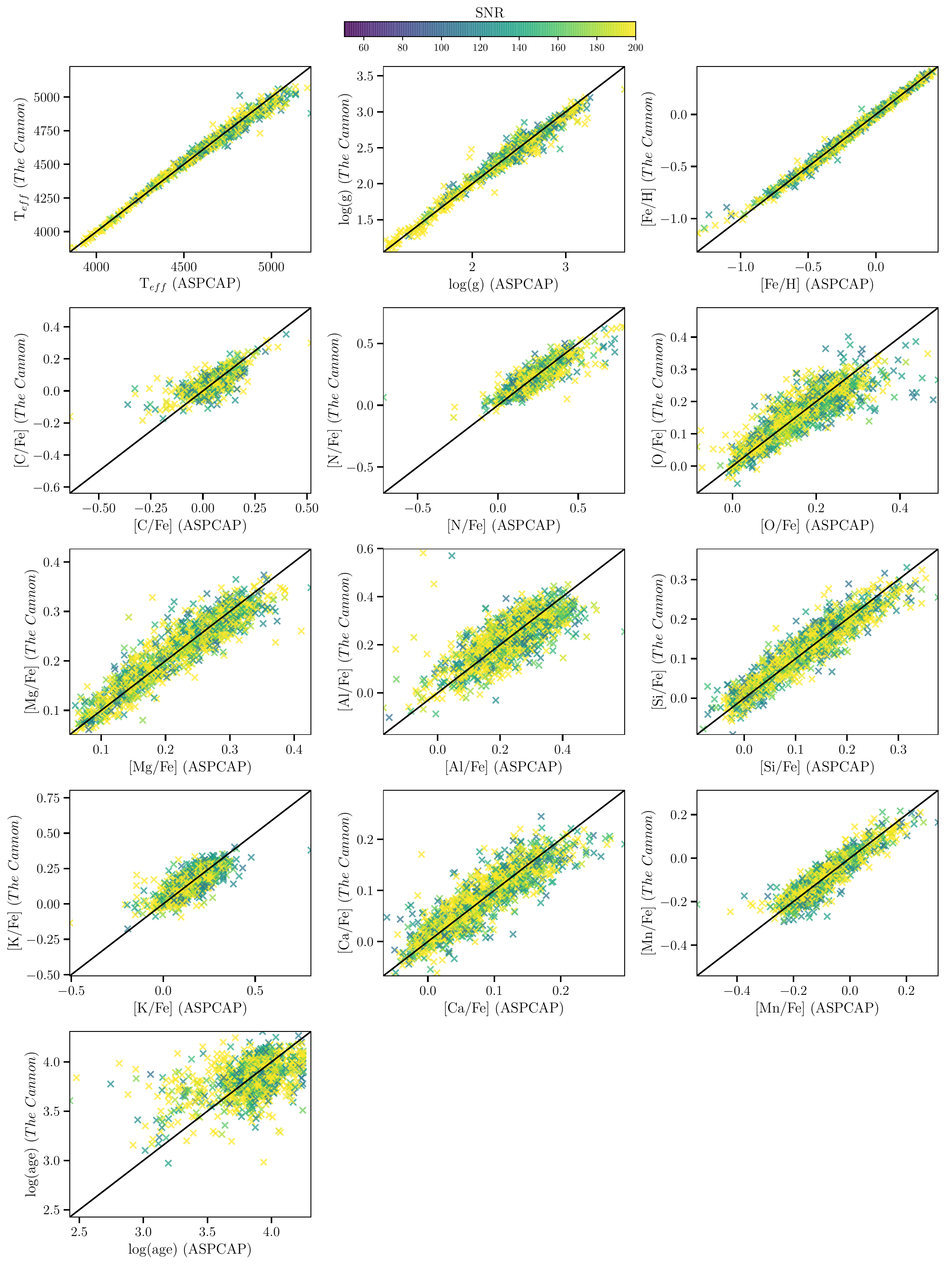}
    \caption{Plots of the reference labels from \aspcap\ vs. the label generated by \textit{The Cannon} for \teff (K), $\log{g}$, [Fe/H], [C/Fe], [N/Fe], [O/Fe], [Mg/Fe], [Al/Fe], [Si/Fe], [K/Fe], [Ca/Fe], [Mn/Fe] (all dex), and $\log({\textrm{age (Myr)}})$. Values from \textit{The Cannon} are result of the 10-fold cross validation of the final 13 label model for the high-$\alpha$ subset of training stars ($N=1255$), with continuum normalized spectra.}
    \label{fig: high-alpha cross-validation plots}
\end{figure*}

\begin{figure*}[!hp]
    \centering
    \includegraphics[width=0.9\textwidth, keepaspectratio]{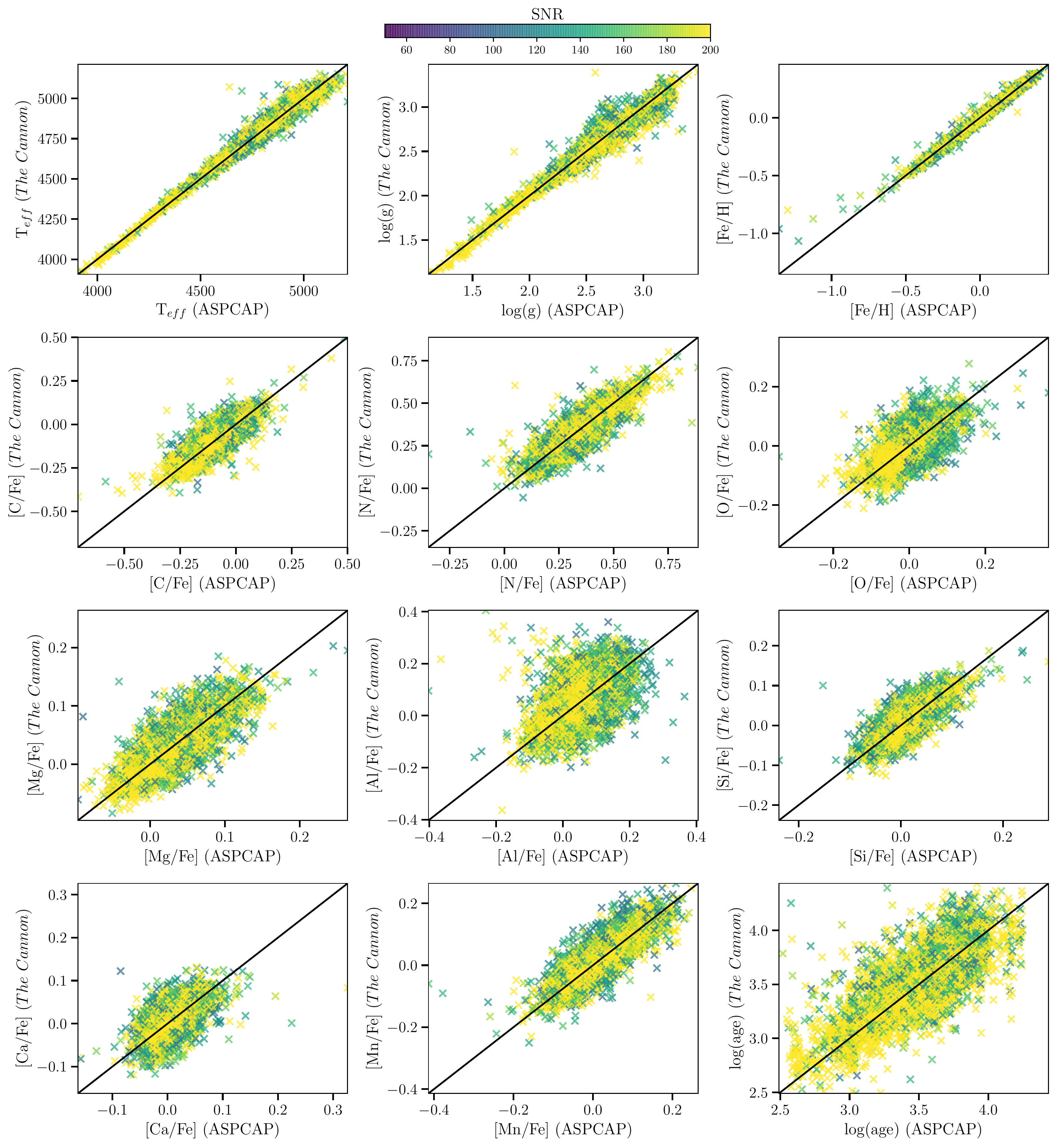}
    \caption{Plots of the reference label from \aspcap vs. the label generated by \textit{The Cannon} for \teff (K), $\log{g}$, [Fe/H], [C/Fe], [N/Fe], [O/Fe], [Mg/Fe], [Al/Fe], [Si/Fe], [Ca/Fe], [Mn/Fe] (all dex) and $\log({\textrm{age (Myr)}})$. Values from \textit{The Cannon} are result of the 10-fold cross validation of the final 12 label model for the low-$\alpha$ subset of training stars ($N=4736$), with continuum normalized spectra.}
    \label{fig: low-alpha cross-validation plots}
\end{figure*}

\section{Methods}
\label{sec: methods}

\begin{figure}[!t]
    \centering
    \includegraphics[scale=0.55, keepaspectratio]{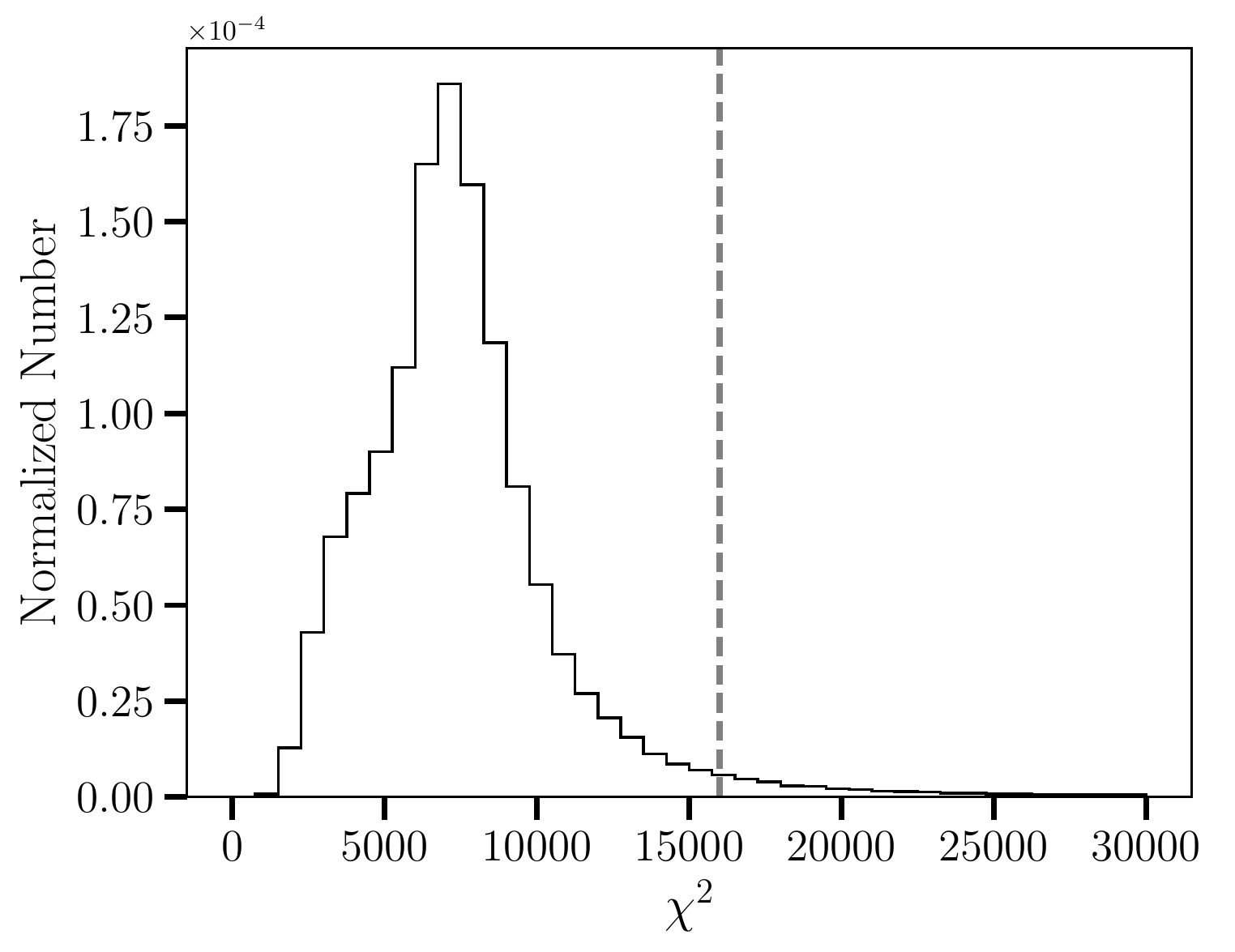}
    \caption{$\chi^2$ distribution of the test set from the model fit compared to the data for each star. We make a conservative cut for our stars we use in our analysis, keeping only stars with $\chi^2<16,000$. This corresponds to a  $\chi^2$ that is  approximately twice the number of pixels as the data. Of total 125,367 stars in the test set, 121,278 made this $\chi^2$ cut.}   
    \label{fig: Chi2}
\end{figure}

\begin{figure}[!t]
    \centering
    \includegraphics[scale=0.54, keepaspectratio]{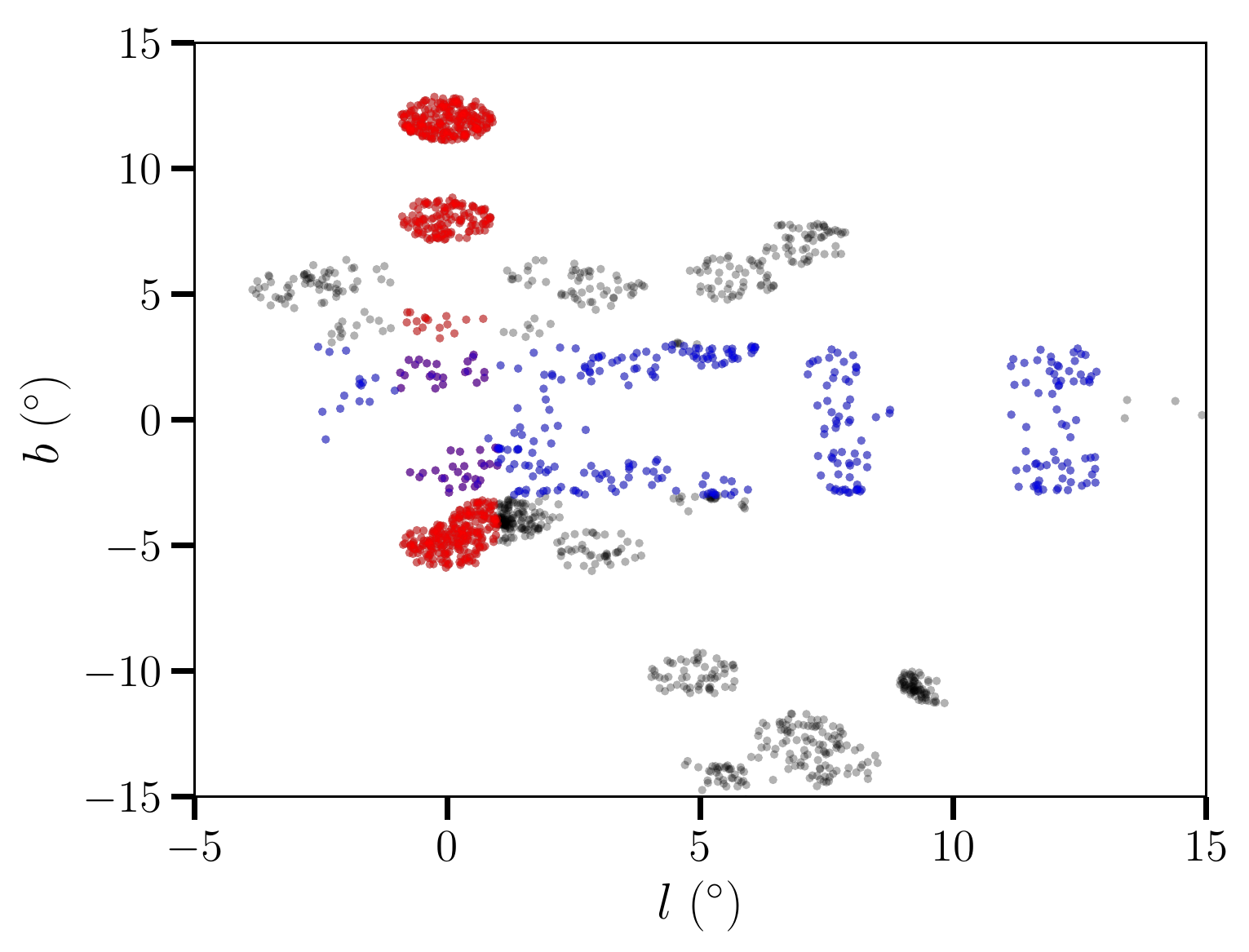}
    \caption{The distribution of observed stars in the bulge region across ($l,b$). We consider the bulge region to be within $(-15^\circ, -15^\circ)<(l, b)<(15^\circ,  15^\circ)$. This yields a total of 1654 stars out of 2567 total stars within \rgal\ $<$ 3.5 kpc of the Galactic center. Red stars indicate those along the minor axis at $l=0^\circ$, that were selected for further analysis in Section \ref{subsec: minor axis analysis}. Blue stars indicate those along the major axis, $b=0^\circ$ , that were selected for further analysis in Section \ref{subsec: major axis analysis}}
    \label{bulge map}
\end{figure}

A correlation exists between abundances and ages that is different between stars with a high abundance of $\alpha$-elements and those with a low abundance of $\alpha$-elements \citep[e.g.][]{Bedell2018}. Therefore, we chose to divide up our sample of stars into a high-$\alpha$ set and low-$\alpha$ set and train two separate models with \textit{The Cannon}. We use the magnesium abundance enhancement, [Mg/Fe],  as a proxy for the overall $\alpha$-element abundance [$\alpha$/Fe]. As this separation does not need to be extremely accurate, a linear cutoff line was visually fit to a plot of [Fe/H] vs. [Mg/Fe] to obtain a cutoff where stars with [Mg/Fe] $>$ 0.132 $\times$ [Fe/H] + 0.111 were considered to be high-$\alpha$. Figure \ref{fig: alpha cutoff} shows this cutoff on a plot of [Fe/H] vs. [Mg/Fe]. This yields 1255 high-$\alpha$ training stars and 4736 low-$\alpha$ training stars.
\par
In \textit{The Cannon}, a second-order polynomial model was trained for each of the high- and low-$\alpha$ training sets. We performed a 10-fold cross validation for each set, training the model on 90\% of the overall training set and testing on the other 10\%, repeating the process 10 times such that every star was tested. This cross validation showed that some of the abundances we had intended to use were not well recovered and the model became unstable with small changes to the training and validation sets using these labels, occasionally failing during one step of the cross validation. Removing labels that caused these failures resulted in a high-$\alpha$ model with 13 labels (\teff, $\log{g}$, [Fe/H], $\log{(\textrm{age})}$, and [X/Fe] for X=C, N, O, Mg, Al, Si, K, Ca, Mn) and a low-$\alpha$ model with 12 labels (the same 13 labels as the high-$\alpha$ case except for [K/Fe]). Figures \ref{fig: high-alpha cross-validation plots} and \ref{fig: low-alpha cross-validation plots} plot the values generated by \textit{The Cannon} compared to the reference values used for training. Figure \ref{fig: high-alpha cross-validation plots} shows this for all 13 high-$\alpha$ labels and Figure \ref{fig: low-alpha cross-validation plots} shows the same for all 12 low-$\alpha$ labels.
\par
The test set of stars were those from the same \apogee\ DR14 catalog as the training set (but not in the Kepler astroseismic catalog with ages).  Since we do have the spectroscopic parameters for the test set from \apogee's pipeline \aspcap, stars whose \teff, $\log{g}$, or [Fe/H] fall outside the range of the training set were also removed. This left 30,029 high-$\alpha$ and 95,338 low-$\alpha$ stars in the test set.  This additional selection of the test set cut helps prevent extrapolation of \textit{The Cannon's} model beyond the data it was trained on. The test spectra were normalized in the same way as the training set.
\par
We used the StarHorse catalog \citep{Queiroz2018} to obtain the distances to the stars. Using these distances, we converted the $(l, b)$ Galactic longitude and latitude coordinates of the stars into $(x, y, z)$ coordinates and also found \rgal, radius from the Galactic center. A total of 265 stars did not have distances available.

\begin{figure}[!t]
    \centering
    \includegraphics[scale=0.55, keepaspectratio]{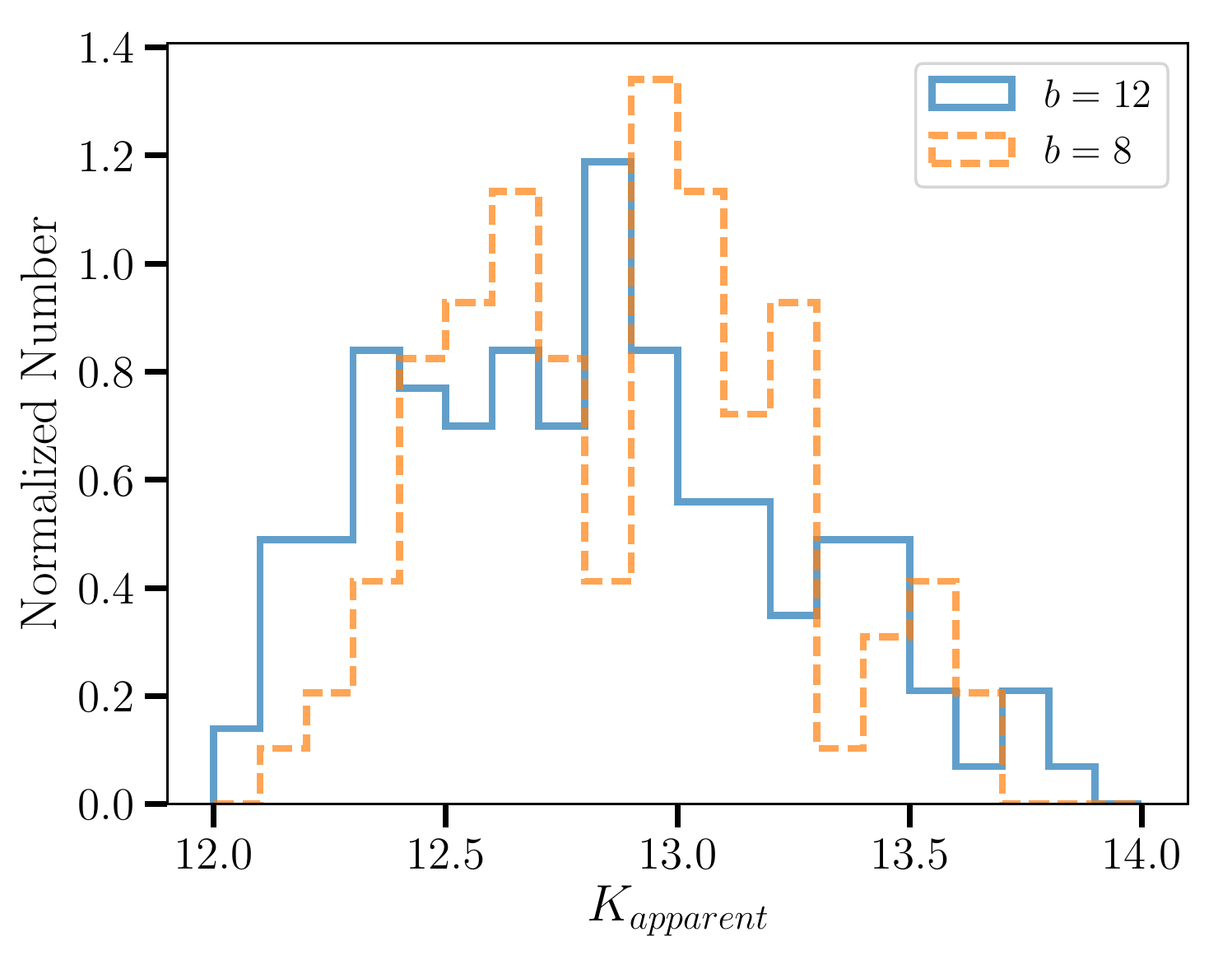}
    \caption{The distribution of $K_{apparent}$ for stars with $2.0<\log{g}<3.0$ in the $(l, b)=(0^\circ, 8^\circ)$ and $(l, b)=(0^\circ, 12^\circ)$ fields. 117 out of total 122 stars in $(l, b)=(0^\circ,  8^\circ)$ and 152 out of total 213 stars in $(l, b)=(0^\circ, 12^\circ)$ satisfy the $log{g}$ cut and are shown in the figure. Total numbers are after the $\chi^2$ cut described in section \ref{sec: results}.}
    \label{fig: Kapparent hist}
\end{figure}

\section{Results} 
\label{sec: results}

\begin{table*}[!t]
\centering
\caption{General description of columns found in table of results, \href{https://drive.google.com/drive/folders/1IfGFbMXd8LrxtGRIrIFJ52A2scF2_gHU?usp=sharing}{available online}. We include results for all 125367 stars in the test set. Note that there is one table with high-$\alpha$ results ($N=30029$) which includes [K/Fe] and one table with low-$\alpha$ results ($N=95338$) which does not include [K/Fe].}
\label{fits file columns}
\begin{tabular}{@{}lll@{}}
\toprule
Column      & Description                                              & Unit          \\ \midrule
ID          & 2MASS-style object name                                  &               \\
Teff        & Effective temperature from \textit{The Cannon}           & K             \\
Teff\_err   & Error in effective temperature                           & K             \\
logg        & Surface gravity from \textit{The Cannon}                 & log(cm s$^{-2}$) \\
logg\_err   & Error in surface gravity                                 & log(cm s$^{-2}$) \\
Fe\_H       & [Fe/H] from \textit{The Cannon}                          & dex           \\
Fe\_H\_err  & Error in [Fe/H]                                          & dex           \\
C\_Fe       & [C/Fe] from \textit{The Cannon}                          & dex           \\
C\_Fe\_err  & Error in [C/Fe]                                          & dex           \\
N\_Fe       & [N/Fe] from \textit{The Cannon}                          & dex           \\
N\_Fe\_err  & Error in [N/Fe]                                          & dex           \\
O\_Fe       & [O/Fe] from \textit{The Cannon}                          & dex           \\
O\_Fe\_err  & Error in [O/Fe]                                          & dex           \\
Mg\_Fe      & [Mg/Fe] from \textit{The Cannon}                         & dex           \\
Mg\_Fe\_err & Error in [Mg/Fe]                                         & dex           \\
Al\_Fe      & [Al/Fe] from \textit{The Cannon}                         & dex           \\
Al\_Fe\_err & Error in [Al/Fe]                                         & dex           \\
Si\_Fe      & [Si/Fe] from \textit{The Cannon}                         & dex           \\
Si\_Fe\_err & Error in [Si/Fe]                                         & dex           \\
K\_Fe       & [K/Fe] from \textit{The Cannon} (only for high-$\alpha$) & dex           \\
K\_Fe\_err  & Error in [K/Fe] (only for high-$\alpha$)                 & dex           \\
Ca\_Fe      & [Ca/Fe] from \textit{The Cannon}                         & dex           \\
Ca\_Fe\_err & Error in [Ca/Fe]                                         & dex           \\
Mn\_Fe      & [Mn/Fe] from \textit{The Cannon}                         & dex           \\
Mn\_Fe\_err & Error in [Mn/Fe]                                         & dex           \\
logage      & Age from \textit{The Cannon}                             & log(Myr)      \\
logage\_err & Error in age                                             & log(Myr)      \\
Chi2        & $\chi^2$ of model fit                                    &               \\ \bottomrule
\end{tabular}
\end{table*}

Table \ref{fits file columns} provides sample columns from the full table of our results, \href{https://drive.google.com/drive/folders/1IfGFbMXd8LrxtGRIrIFJ52A2scF2_gHU?usp=sharing}{available online}. For all 125,367 test stars, we include \teff, \logg, [Fe/H], abundances, and ages as well as all associated label errors and the $\chi^2$ for the generated model compared to the data given the errors for 8575 pixels. Although we include all results for all 125,367 stars, we make a general $\chi^2$ cut, keeping only stars with $\chi^2<16,000$, which leaves 121,278 stars for analysis. The $\chi^2$ distribution of the full set of
125,367 stars and the $\chi^2$ cutoff is shown in Figure \ref{fig: Chi2}. Additionally, except in section \ref{sec: compare to disk}, we make a distance cut and keep only stars within 3.5 kpc of the Galactic center for our bulge analysis, leaving 2567 stars.
\par
Since the \apogee\ \teff, $\log{g}$, [Fe/H], and abundances are all available for the test set, we are able to not only refine our test set to stars within the stellar parameter range of our training set, but also compare our results from \textit{The Cannon} to the \aspcap\ labels. We examine the \apogee\ stellar parameter and abundance labels vs. the labels from \textit{The Cannon} for test stars within the ranges in the training set for all the stellar parameter and abundance labels. For \teff, we notice a set of stars which are dissimilar between the two sets of labels, where \textit{The Cannon} found very high temperatures compared to the \apogee\ value. We checked on all \apogee\ flags and could not see any issues with these stars. These stars actually make up less than 0.5\% of the total test set however and we do not exclude them from our catalog. 

\begin{figure*}[!t]
    \centering
    \includegraphics[width=\textwidth, keepaspectratio]{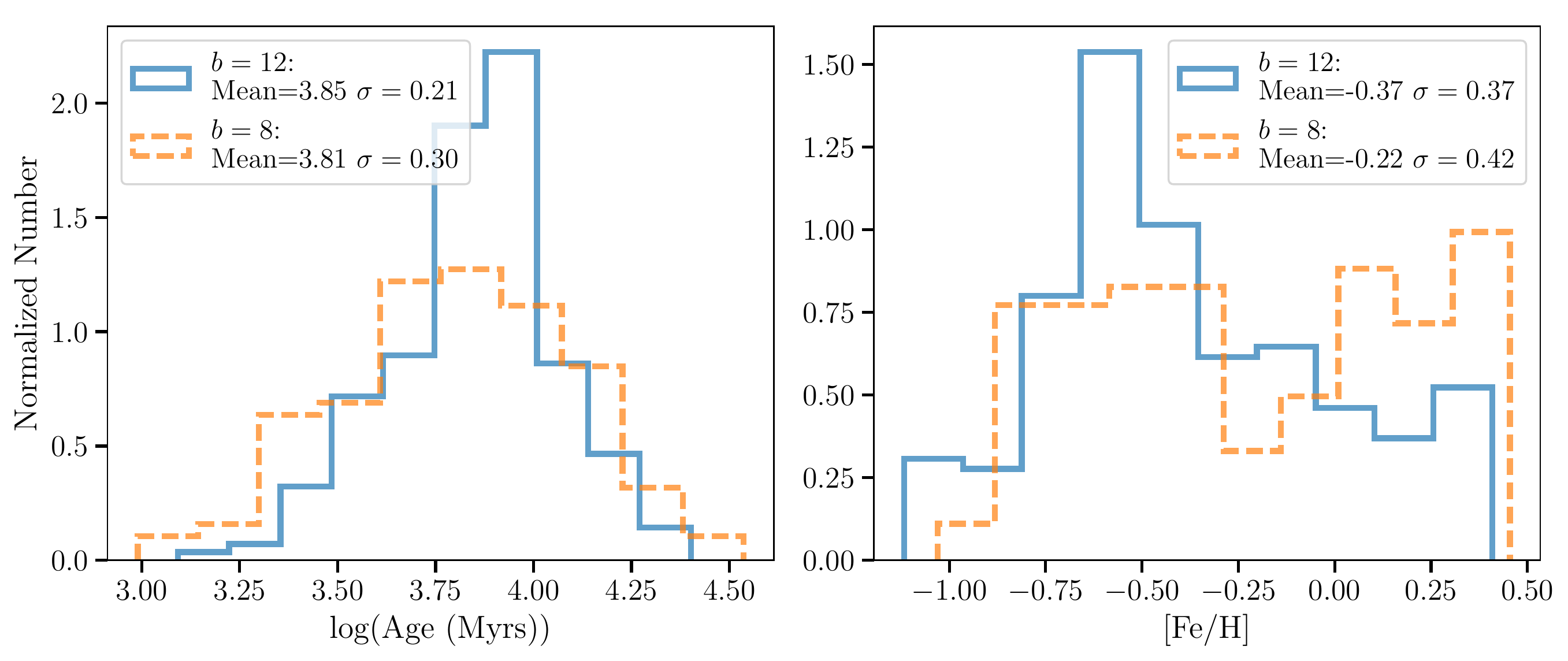}
    \includegraphics[width=0.55\textwidth, keepaspectratio]{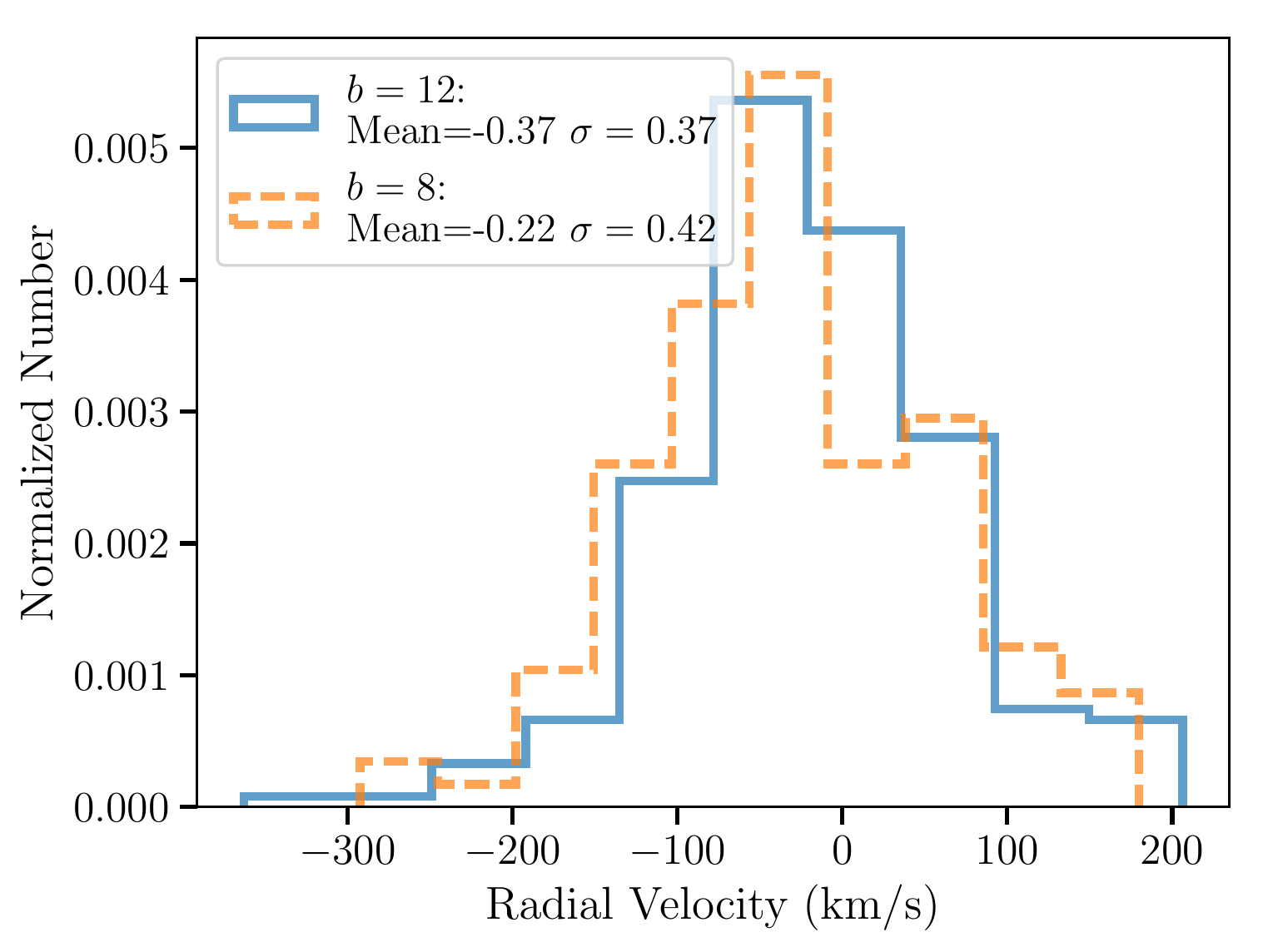}
    \caption{Top left: histograms of the age distributions for the $(l, b)=(0,8)$ and $(l, b)=(0,12)$ fields. Top right: histograms of the metallicity distributions as derived by \textit{The Cannon} for the same fields. Bottom: radial velocity distributions of the fields. The $(l, b)=(0,8)$ field contains 122 stars while the $(l, b)=(0,12)$ field contains 213 stars.}
    \label{fig: b8 and b12 histograms}
\end{figure*}

\begin{figure*}[!ht]
    \centering
    \includegraphics[width=\textwidth, keepaspectratio]{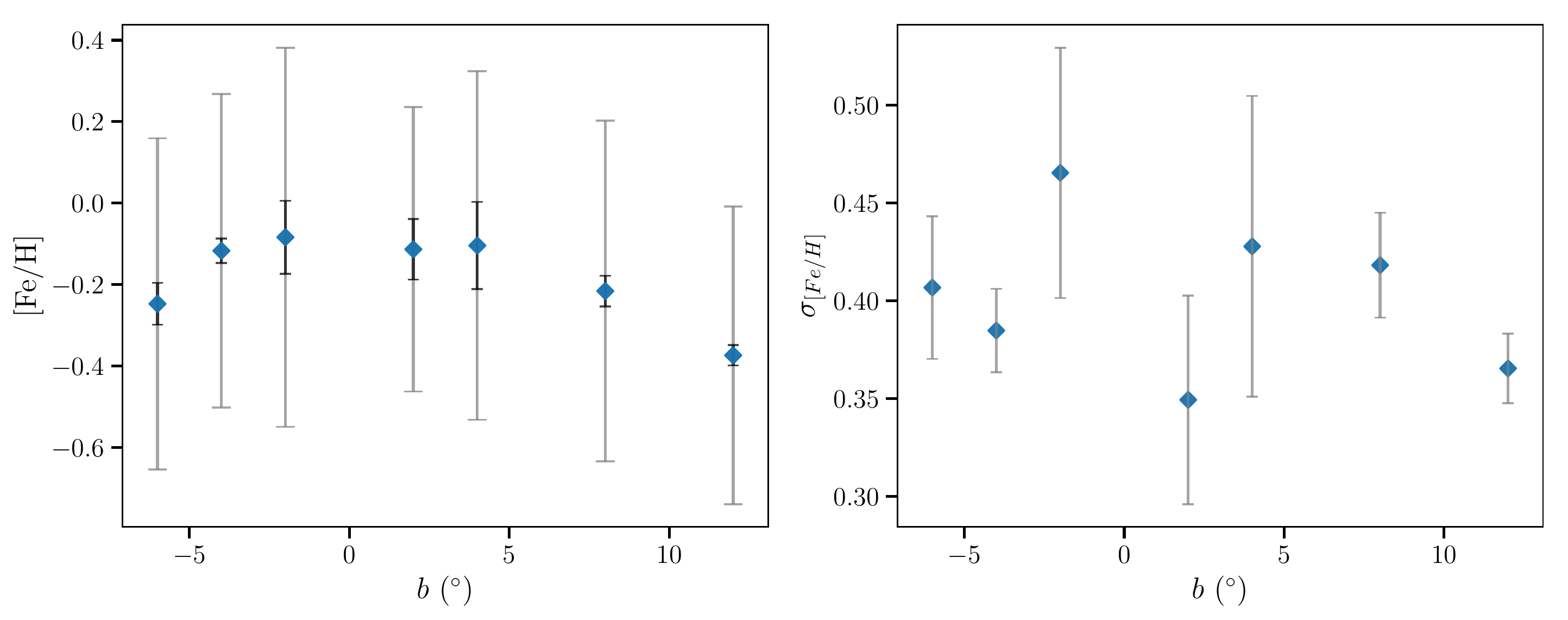}
    \caption{Left, the mean [Fe/H] as a function of latitude along the minor axis at $l$ = 0$^{\circ}$. The light error bars show the 1$\sigma$ standard deviation around the mean [Fe/H] and the (much smaller) dark error bars show the confidence on the mean measurement. Right: the 1$\sigma$ standard deviation of the mean in each field with error bars indicating the confidence on the measurement. There are 627 stars in each figure.}
    \label{fig: [Fe/H] along l=0}
\end{figure*}

\begin{figure*}[!ht]
    \centering
    \includegraphics[width=\textwidth, keepaspectratio]{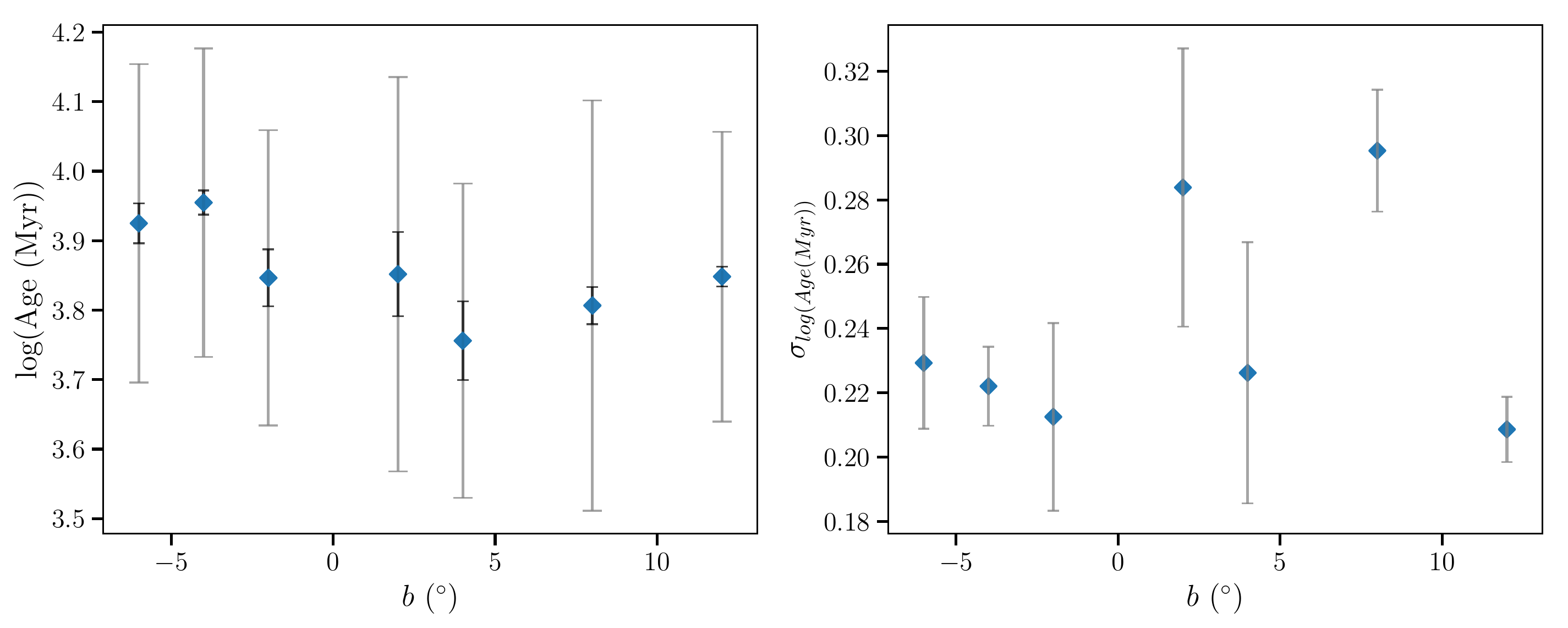}
    \caption{Left: the mean age as a function of latitude along the minor axis at $l$ = 0$^{\circ}$. The light error bars show the 1$\sigma$ standard deviation around the mean log(age) and the (much smaller) dark error bars show the confidence on the mean measurement. Right: the 1$\sigma$ standard deviation of the mean in each field with error bars indicating the confidence on the measurement. There are 627 stars in each figure.}
    \label{fig: ages along l=0}
\end{figure*}

\begin{figure*}[!ht]
    \centering
    \includegraphics[width=\textwidth, keepaspectratio]{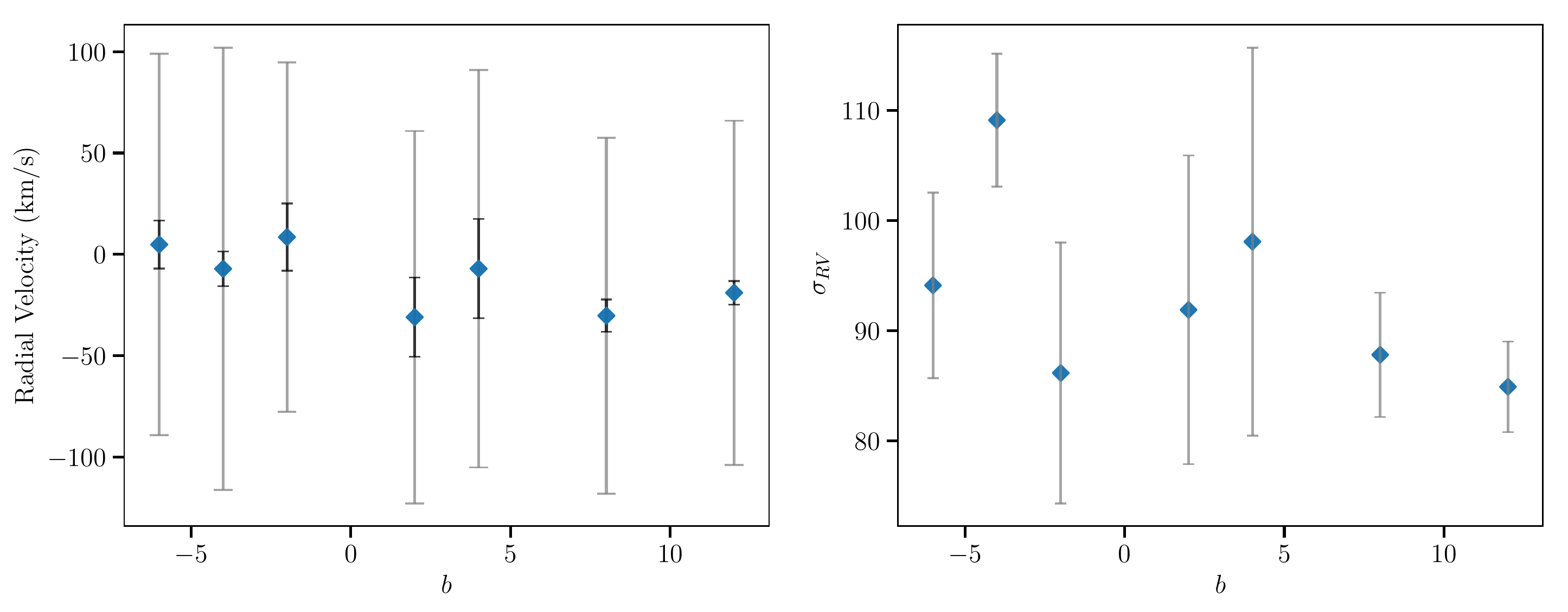}
    \caption{Left: the mean radial velocity as a function of latitude along the minor axis at $l$ = 0$^{\circ}$. The light error bars show the 1$\sigma$ standard deviation around the mean log(age) and the dark (much smaller) error bars show the confidence on the mean measurement. Right: the 1$\sigma$ standard deviation of the mean in each field with  error bars indicating the confidence on the measurement. There are 627 stars in each figure.}
    \label{fig: RV along l=0}
\end{figure*}

\begin{figure*}[!ht]
    \centering
    \includegraphics[width=\textwidth, keepaspectratio]{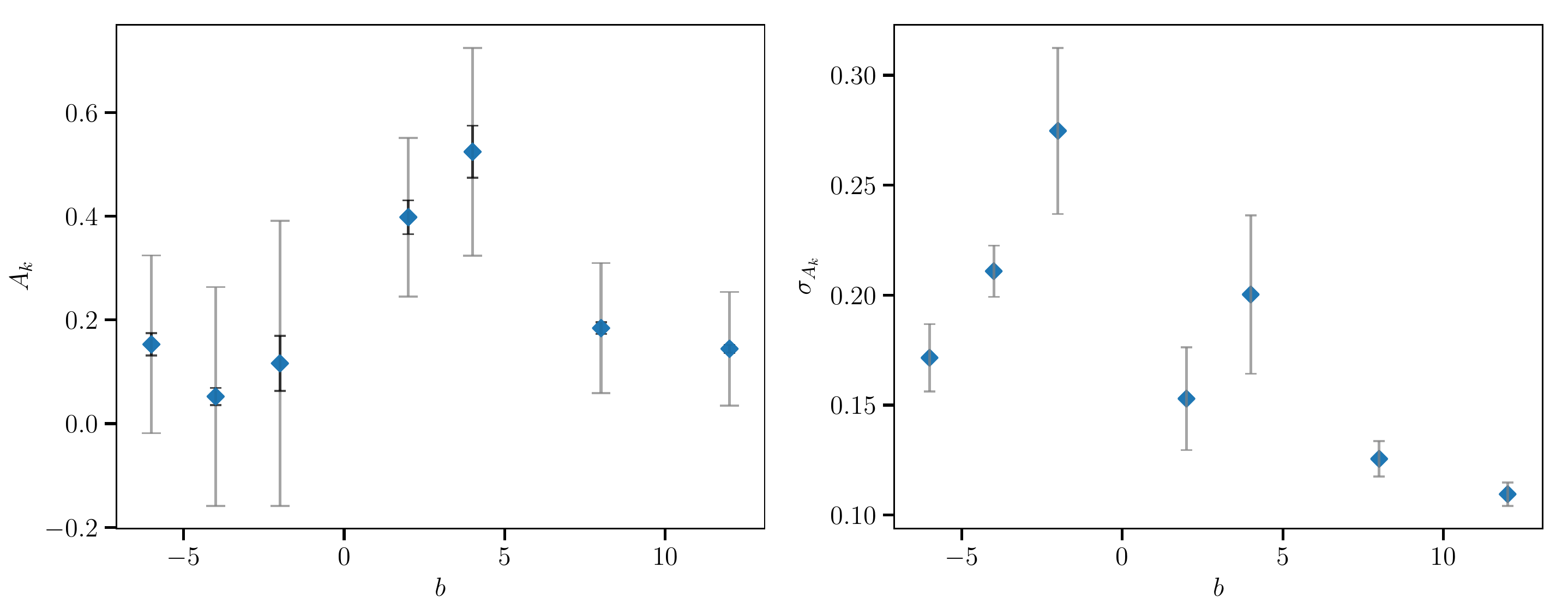}
    \caption{Left: the mean $K$-band extinction $A_k$ as a function of latitude along the minor axis at $l$ = 0$^{\circ}$. The light error bars show the 1$\sigma$ standard deviation around the mean log(age) and the (much smaller) dark error bars show the confidence on the mean measurement. Right: the 1$\sigma$ standard deviation of the mean in each field with error bars indicating the confidence on the measurement. There are 627 stars in each figure.}
    \label{fig: Ak along l=0}
\end{figure*}

Unless otherwise noted, for these analyses, the high- and low-$\alpha$ stars were both used together in examining our bulge ages, metallicities and abundance distributions. There were a few stars that we obtained labels for using \textit{The Cannon} that did not have distances in the StarHorse catalog; 52 of the stars we kept after the $\chi^2$ cut were removed from our analysis for this reason.
\par
We select only stars within \rgal\ $<$ 3.5 kpc for our analysis, which leaves us with a total of 2567 stars. From these 2567 stars, we further select stars within the bulge region that we define to be $(-15^\circ, -15^\circ)<(l, b)<(15^\circ, 15^\circ)$, leaving 1654 bulge stars for our analysis. These stars have typical distance errors of 0.86 kpc. Monte Carlo modeling of 1000 samples shows that these distance errors induce an uncertainty of 5\% in stars we select with \rgal\ $<$ 3.5 kpc. There is less than 2\% uncertainty on all following mean and dispersion measurements of age and metallicity due to this uncertainty in stars we select. Using this set of 1654 stars, we want to examine how stellar ages, metallicities and abundances change over the extent of the bulge and with radius in the inner region. With DR14 we have a limited set of fields to examine within 3.5 kpc of the Galactic center. These fields are plotted in Figure \ref{bulge map}.  We therefore break up our examination into the following: (i) along the minor axis, (ii) along the major axis, (iii) across the entire bulge, and (iv) comparing the bulge to the disk. Our results are detailed in the following subsections.

\subsection{Minor-axis Analysis}
\label{subsec: minor axis analysis}

We first focus on the $(l, b)=(0^\circ, 8^\circ)$ and $(l, b)=(0^\circ, 12^\circ)$ fields. These two fields had special targeting applied to obtain red clump stars \citep{Zasowski2017}, so we consider them in isolation. We only include stars within 3.5 kpc of the Galactic center in our analysis of these fields. This yields 122 stars in the $(l, b)=(0^\circ, 8^\circ)$ field and 213 stars in the $(l, b)=(0^\circ, 12^\circ)$ field.  
\par
The majority of the stars in both fields have $\log{g}$ between 2.0 and 3.0 dex. We select the stars that fall within this $\log{g}$ range only, which further excludes the few stars which are likely not red clump \citep[e.g.][]{Ness2013a}. For the remaining 117 and 152 stars in the fields at $b=8^\circ$ and $b=12^\circ$ respectively, we plot the dereddened apparent magnitude in the $K$ band in Figure \ref{fig: Kapparent hist}, which for the red clump is a proxy for distance along the line of sight. We calculate the dereddened apparent magnitude by subtracting with Wide-field Infrared Survey Explorer $K$-band extinction from the Two Micron All Sky Survey magnitude; both quantities are given in the \aspcap\ data file. Doing so reveals a ``gap" in apparent magnitudes in the  $b=8^\circ$ field that disappears by the $b=12^\circ$ field. This ``gap" is presumably tracing the X-shaped structure in the Milky Way bulge, and this finding shows that the line-of-sight observations do not traverse the arms of the X-shape at Galactic latitudes of $b=12^\circ$, or 1.7 kpc above the plane in the center of the bulge. This is also consistent with the \cite{Ness2016b} image of the bulge, which shows the X-shape extending to slightly below $b=\pm10^\circ$. This also indicates that the field at $b=8^\circ$ is presumably tracing the ends of the X-shape structure. 

\begin{figure}[!h]
    \centering
    \includegraphics[width=0.47\textwidth, keepaspectratio]{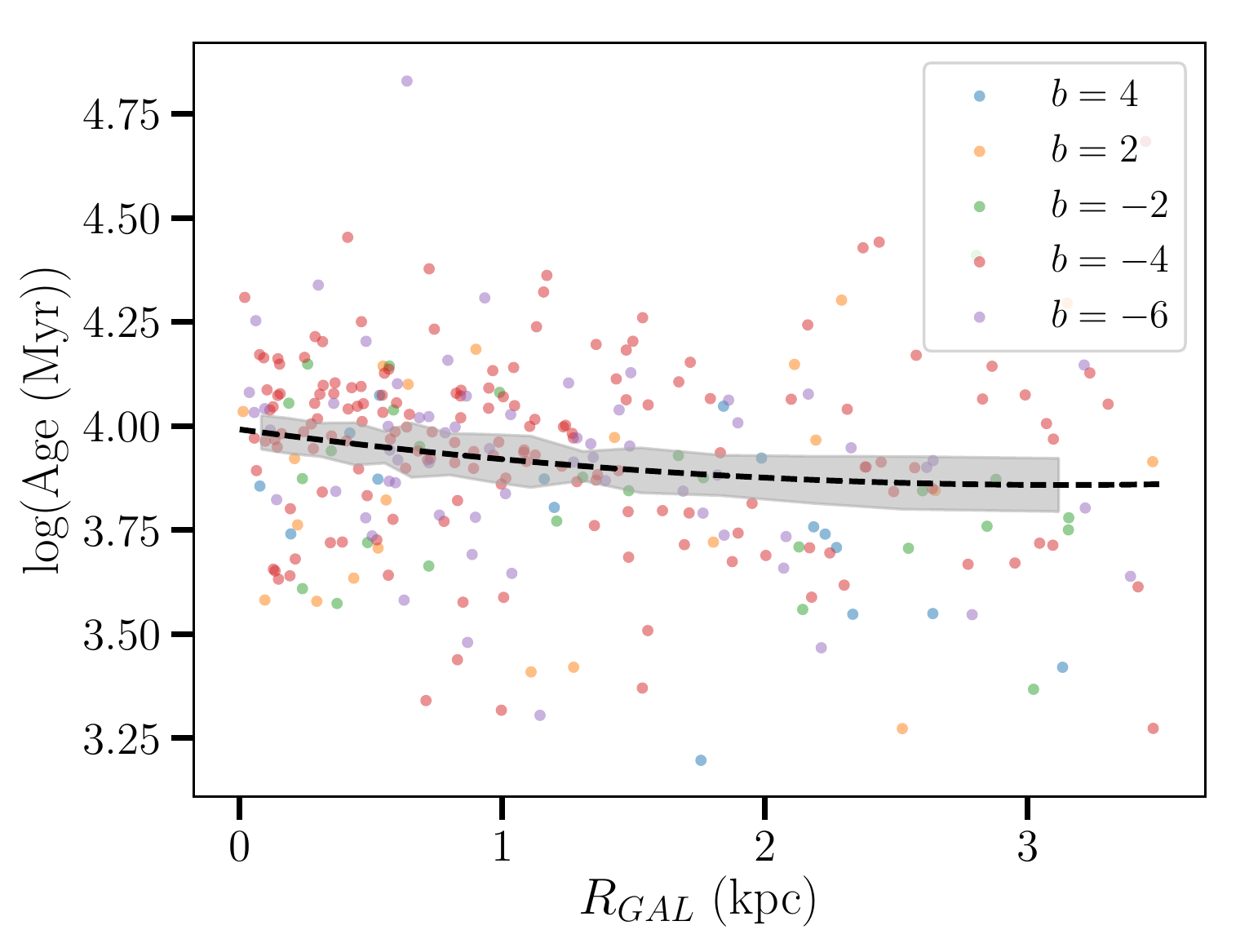}
    \includegraphics[width=0.47\textwidth, keepaspectratio]{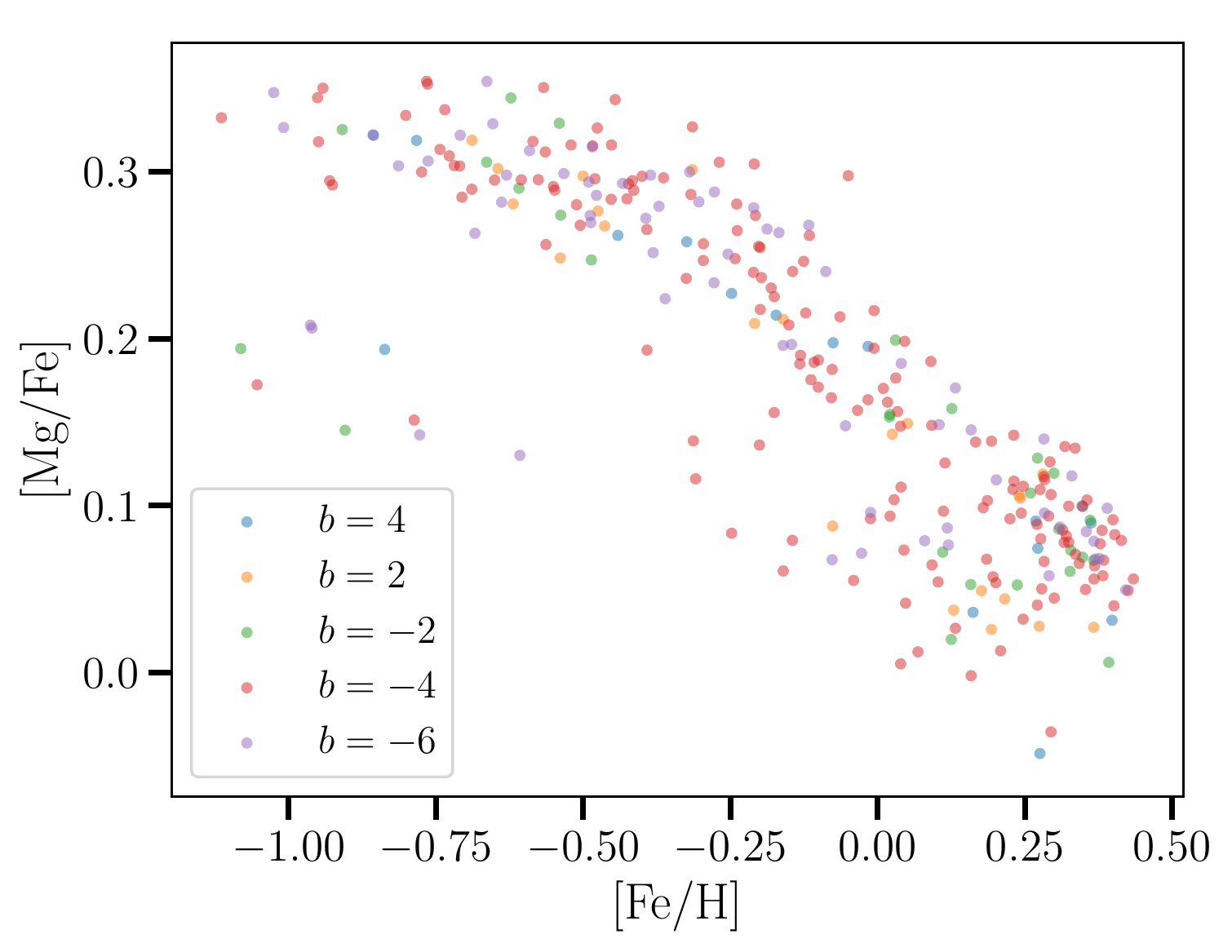}
    \caption{At top, age as a function of distance from the Galactic center for fields centered at $(l, b) = (0^{\circ}, -6^{\circ})$, $(0^{\circ}, -4^{\circ})$, $(0^{\circ}, -2^{\circ})$, $(0^{\circ}, 2^{\circ})$, $(0^{\circ}, 4^{\circ})$ ($N=$ 291). A second-order polynomial (black dashed line) has been fit to all the stars, and standard error, calculated using 15 bins, is depicted in gray. At bottom, [Fe/H] vs. [Mg/Fe] for the same fields.}
    \label{fig: low lat Rgal vs age}
\end{figure}

\begin{figure}[!h]
    \centering
    \includegraphics[width=0.47\textwidth, keepaspectratio]{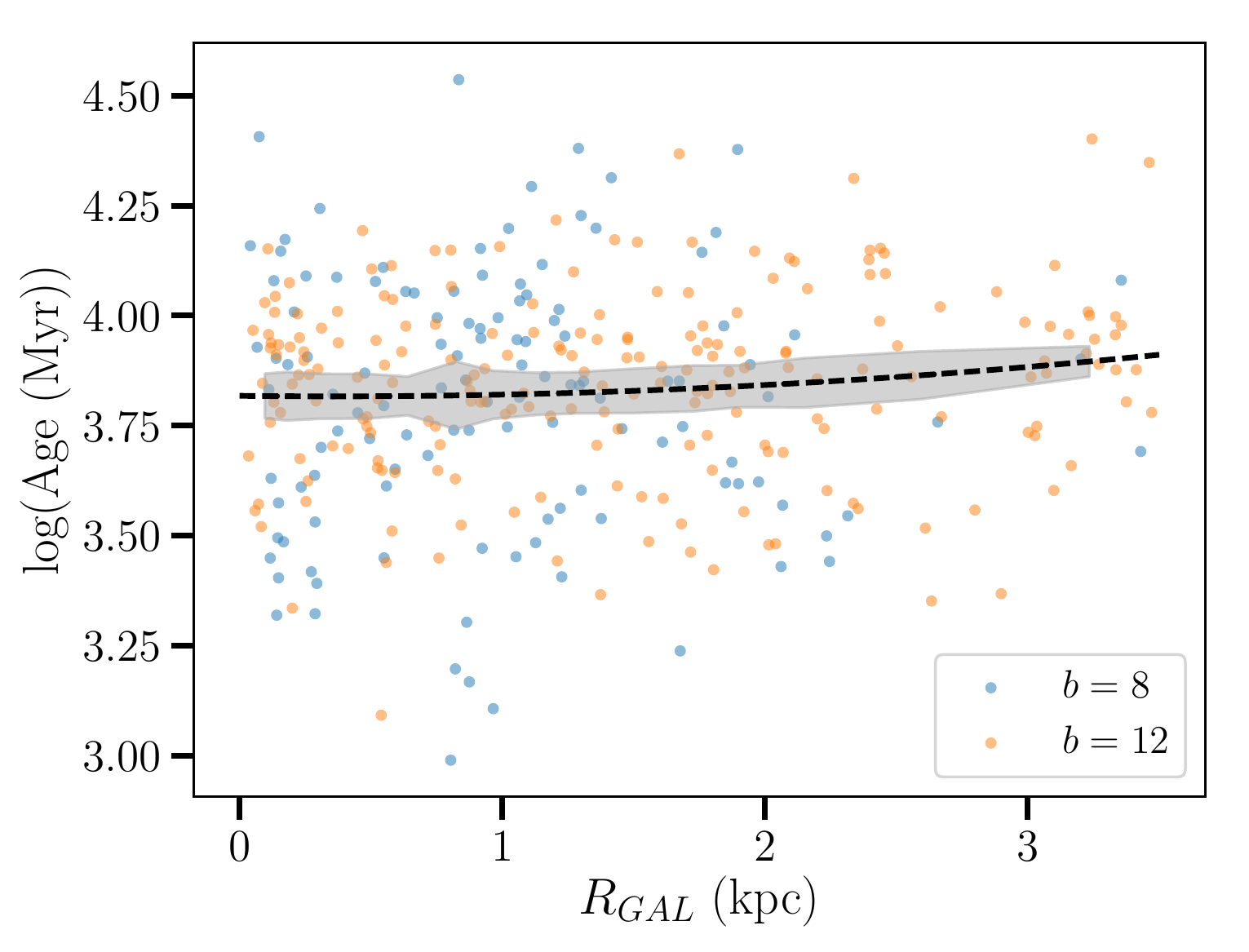}
    \includegraphics[width=0.47\textwidth, keepaspectratio]{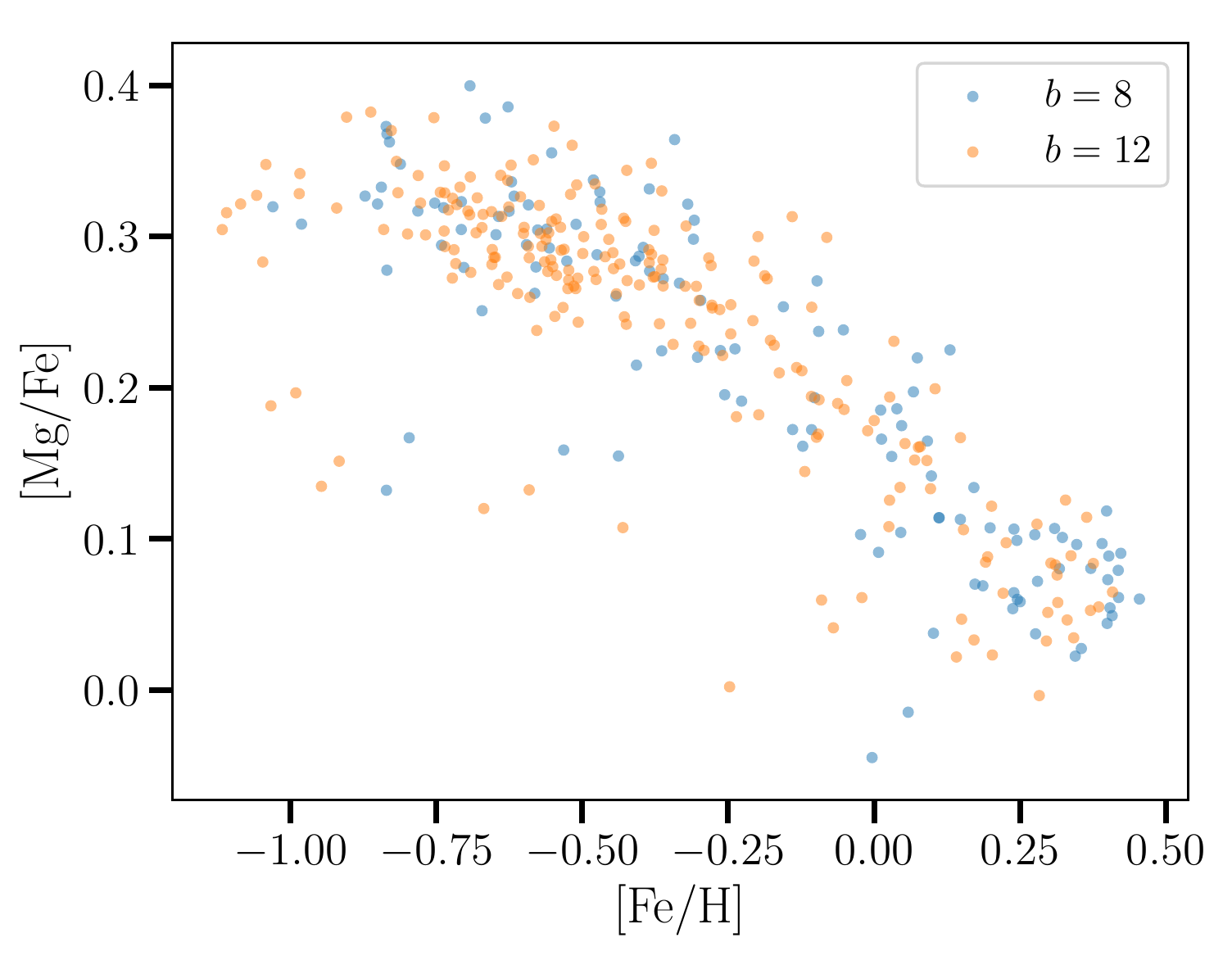}
    \caption{At top, age as a function of distance from the Galactic center for fields centered at $(l, b) = (0^{\circ}, 8^{\circ})$, $(0^{\circ}, 12^{\circ})$ ($N=335$). A second-order polynomial (black dashed line) has been fit to all the stars, and standard error, calculated using 15 bins, is depicted in gray. At bottom, [Fe/H] vs. [Mg/Fe] for the same fields.}
    \label{fig: high lat Rgal vs age}
\end{figure}

The left-hand panel of Figure \ref{fig: b8 and b12 histograms} shows the age distributions for these two fields. The two fields seem to have a similar mean age of $\tau \approx$ 7 Gyr. However, the lower, $(l, b)=(0^\circ, 8^\circ)$, field has an $\approx$40 percent larger age dispersion. The right-hand panel of Figure \ref{fig: b8 and b12 histograms} shows the [Fe/H] distribution of these two fields. The metallicity distribution of the two fields is more strikingly dissimilar between the fields. The $(l, b)=(0^\circ, 8^\circ)$ field has two peaks around [Fe/H] = 0.25 dex and [Fe/H] = 0.6 dex and has a larger fraction of metal-rich stars compared to the higher-latitude field. The field at $(l, b)=(0^\circ, 12^\circ)$ has a significantly larger fraction of metal poor stars around [Fe/H] = -0.6 dex. However, we did not observe any clear trends in the metallicity dependence of the X-shape; this could be due to the sample size being too small for a full investigation. We also find for both fields that stars of fixed age have a broad and similar range in [Fe/H].

\begin{figure*}[!ht]
    \centering
    \includegraphics[width=\textwidth, keepaspectratio]{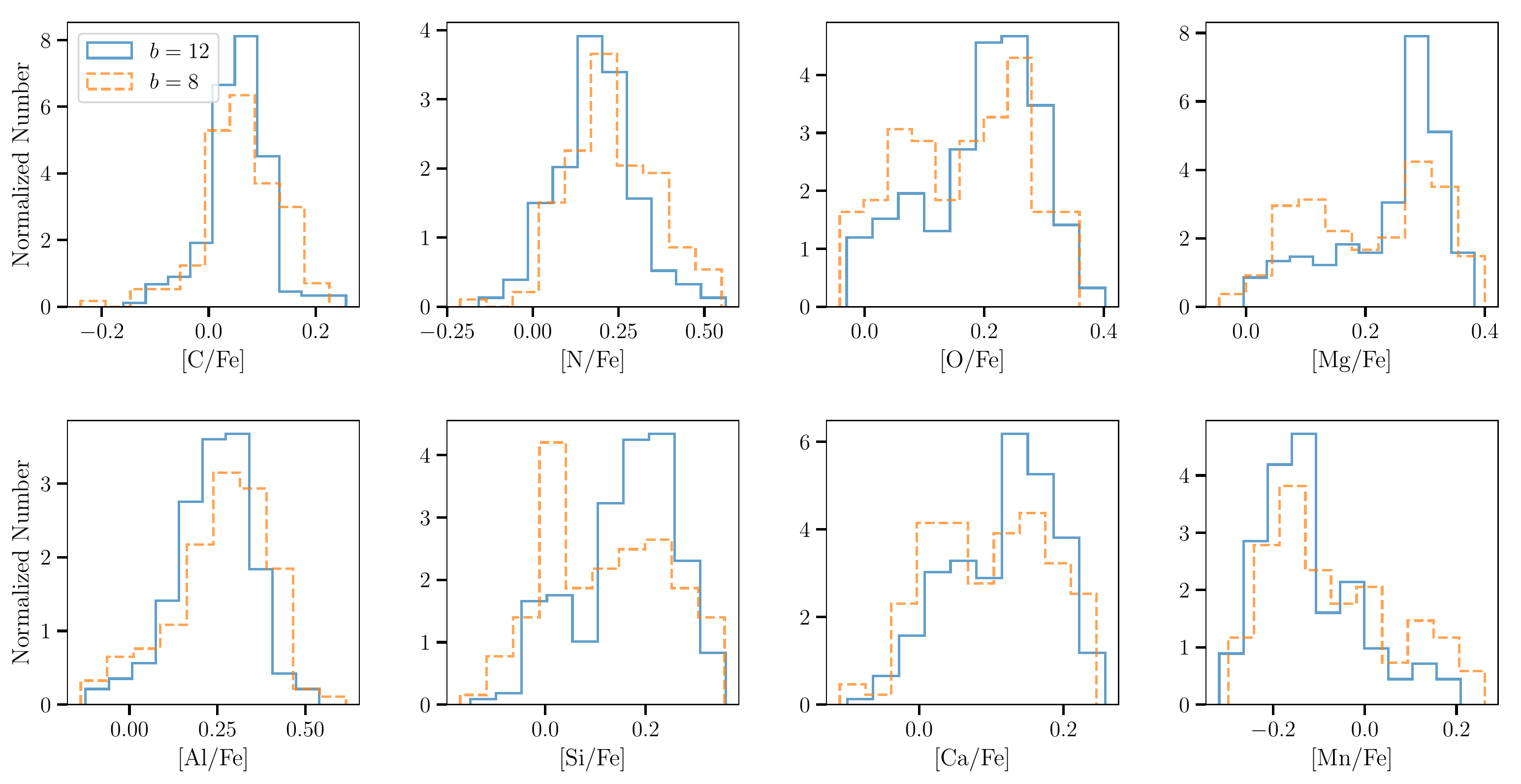}
    \caption{Distributions of elemental abundances from \textit{The Cannon} using unnormalized spectra of the $(l, b) = (0^{\circ},8^{\circ})$  and $(0^{\circ},12^{\circ})$ fields. We do not show [K/Fe] since it was only included in the high-$\alpha$ model. }
    \label{fig: element hists}
\end{figure*}

We now examine the ages and metallicities of stars along the minor axis, $l=0^\circ$, for all $b$ within \rgal\ $<$ 3.5 kpc of the Galactic center.  The previously discussed $(l, b)=(0^\circ, 8^\circ)$ and $(l, b)=(0^\circ, 12^\circ)$ fields each extend about 1$^\circ$ radially from their centers. We consider the stars by a set of spatial fields. The $(l, b)=(0^\circ, 8^\circ)$ field encompasses stars between $(-1^\circ, 7^\circ)<(l, b)<(1^\circ, 9^\circ)$; the $(l, b)=(0^\circ, 12^\circ)$ field encompasses stars between $(-1^\circ, 11^\circ)<(l, b)<(1^\circ, 13^\circ)$. Following this, we select all remaining stars with $-1^\circ<l<1^\circ$ and bin them every 2$^\circ$, resulting in six additional groups of $l=0^\circ$ stars. However, the group centered at $(l, b)=(0^\circ, 0^\circ)$ contained far fewer stars than the others (seven stars), so we exclude it from further analysis. This results in a total of 627 stars for our minor-axis analysis.
\par
The left panel of Figure \ref{fig: [Fe/H] along l=0} shows the mean metallicities for the six fields along $l=0^\circ$, as described. The sampling error, $\nicefrac{\sigma}{\sqrt{N}}$, where $N$ is the number of stars in the field, is also plotted as a measure of confidence in the mean measurement. The right panel of Figure \ref{fig: [Fe/H] along l=0} shows the standard deviation of metallicities of the six fields along $l=0^\circ$, and we also plot their sampling errors, $\nicefrac{\sigma}{\sqrt{2N-1}}$. Figures \ref{fig: ages along l=0}, \ref{fig: RV along l=0}, and \ref{fig: Ak along l=0}  show the same for the ages, radial velocity, and $K$-band extinction, respectively. We include velocity and $K$-band extinction as checks for correlations or systematics (i.e. that could impact the distance distribution of the stars) which may affect our age and metallicity results. 
\par
As seen in the left panel of Figure \ref{fig: [Fe/H] along l=0}, we find that along $l=0^\circ$, the metallicity decreases from $b=0^\circ$, both above and below the plane. In Figure \ref{fig: ages along l=0}, for age along $l=0^\circ$, unlike for metallicity, we find a slight asymmetry. The stars at negative $b$ seem to be older on average than those at positive $b$. This asymmetry is likely related to extinction effects: Figure \ref{fig: Ak along l=0} shows that the negative $b$ fields have lower mean $A_k$ and greater scatter in $A_k$ than the positive $b$ fields. Compared to those at positive $b$, the negative $b$ fields have generally more stars close to the Galactic center, so the asymmetry may also be related to the distance distribution. We also find a very slight asymmetry in the same direction, higher at negative $b$ and lower at positive $b$, in the average radial velocity, shown in Figure \ref{fig: RV along l=0}. The age and [Fe/H] dispersion are variable along the minor axis and in general uncorrelated. The highest-latitude field however at $(l, b) = (0^\circ, 12^\circ)$ has both the lowest age and [Fe/H] dispersion. 
\par
We now want to examine trends of age with Galactic radius for the stars in these fields. We consider the high and low latitudes separately. In the top panel of Figure \ref{fig: low lat Rgal vs age}, we show age as a function of Galactic radius for our lower latitude fields of $(l, b) = (0^\circ, -6^\circ)$, $(0^\circ, -4^\circ)$, $(0^\circ, -2^\circ)$, $(0^\circ, 2^\circ)$, $(0^\circ, 4^\circ)$. We show the same for our high latitude fields of $(l, b) = (0^\circ, 8^\circ)$ and $(0^\circ, 12^\circ)$ in the top panel of Figure \ref{fig: high lat Rgal vs age}.
\par
We find slightly different age-radius trends in the high-latitude combined fields compared to the lower-latitude combined fields. We fit a second-order polynomial to all the stars in each combined field. Comparing the high and lower latitudes, we find that near the Galactic center, the higher-latitude fields are on average younger than those of the lower-latitude fields. Also, while at higher latitudes we generally find slightly older stars further away from the Galactic center, the opposite is true for the lower-latitude fields. The slope of the age-radius gradient is also shallower for the $(l, b) = (0^\circ, 8^\circ)$ and $(0^\circ, 12^\circ)$ fields.

\subsubsection{Element Abundance Trends along the Minor Axis}

In addition to age and metallicity, we also examine the distributions of the other elements in the $(l, b) = (0^\circ, 8^{\circ})$ compared to the $(l, b) = (0^\circ, 12^\circ)$ fields. We plot these distributions in Figure \ref{fig: element hists}. Interestingly, the distributions are the same for some of the elements but quite different for others. For the elements that are different between the fields, particularly [O/Fe], [Mg/Fe], [Si/Fe], and [Ca/Fe], the distributions differ in the same way: the $(l, b) = (0^\circ, 8^\circ)$ field has a larger proportion of less enriched stars.  That is, there are a significantly larger fraction of low-$\alpha$ stars at lower latitudes in the inner Galaxy compared to at higher latitudes, similarly to near the Sun \citep[e.g.][]{Bensby2012, Hayden2015, Nidever2014}.

\subsection{Age Trends along the Major Axis}
\label{subsec: major axis analysis}

\begin{figure*}[!ht]
    \centering
    \includegraphics[width=\textwidth, keepaspectratio]{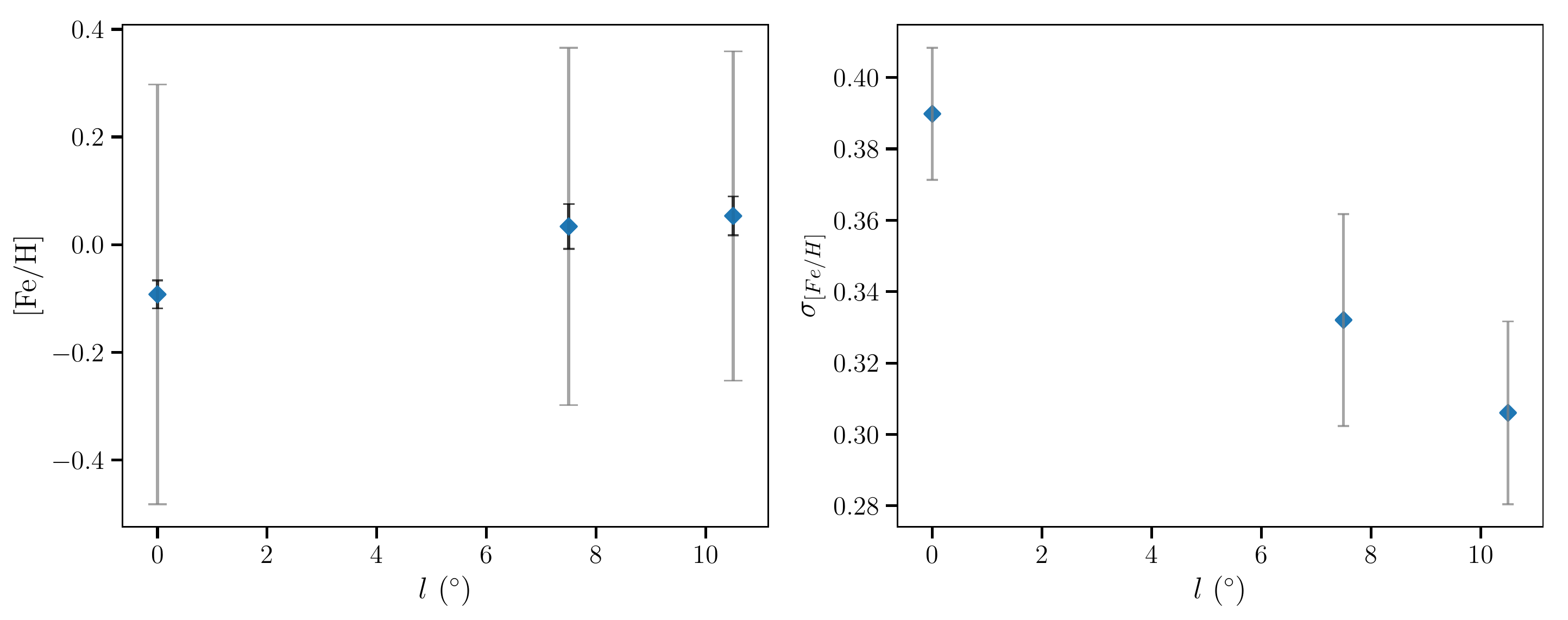}
    \caption{Left: the mean [Fe/H] as a function of longitude along the major axis, at $b$ = 0$^{\circ}$. The light error bars show the 1$\sigma$ standard deviation around the mean [Fe/H] and the (smaller) dark error bars show the confidence on the mean measurement. Right: the 1$\sigma$ standard deviation of the mean in each field with error bars indicating the confidence on the measurement. There are 336 stars that comprise this figure.}
    \label{fig: [Fe/H] along b=0}
\end{figure*}

\begin{figure*}[!ht]
    \centering
    \includegraphics[width=\textwidth, keepaspectratio]{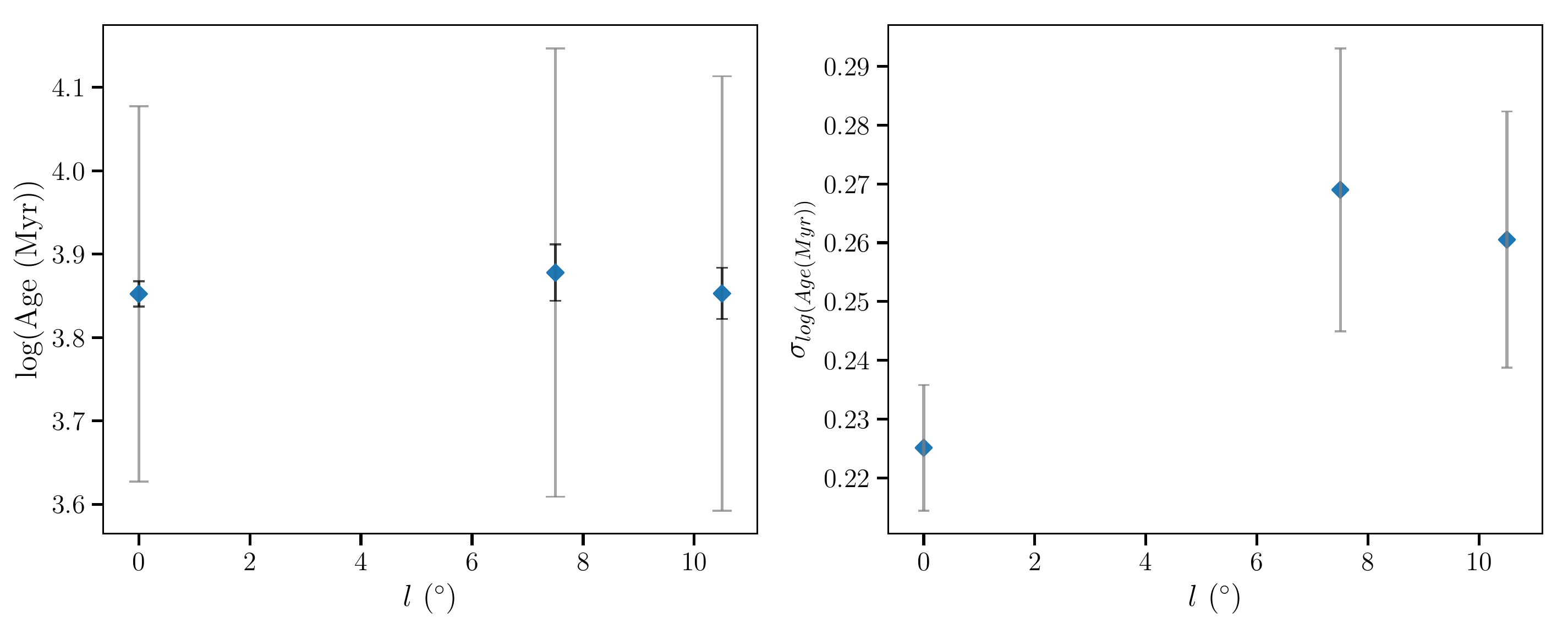}
    \caption{Left: the mean age as a function of longitude along the major axis, at $b$ = 0$^{\circ}$. The light error bars show the 1$\sigma$ standard deviation around the mean age and the (smaller) dark error bars show the confidence on the mean measurement. Right: the 1$\sigma$ standard deviation of the mean in each field with error bars indicating the confidence on the measurement. There are 336 stars that comprise this figure.}
    \label{fig: ages along b=0}
\end{figure*}

\begin{figure*}[!ht]
    \centering
    \includegraphics[width=\textwidth, keepaspectratio]{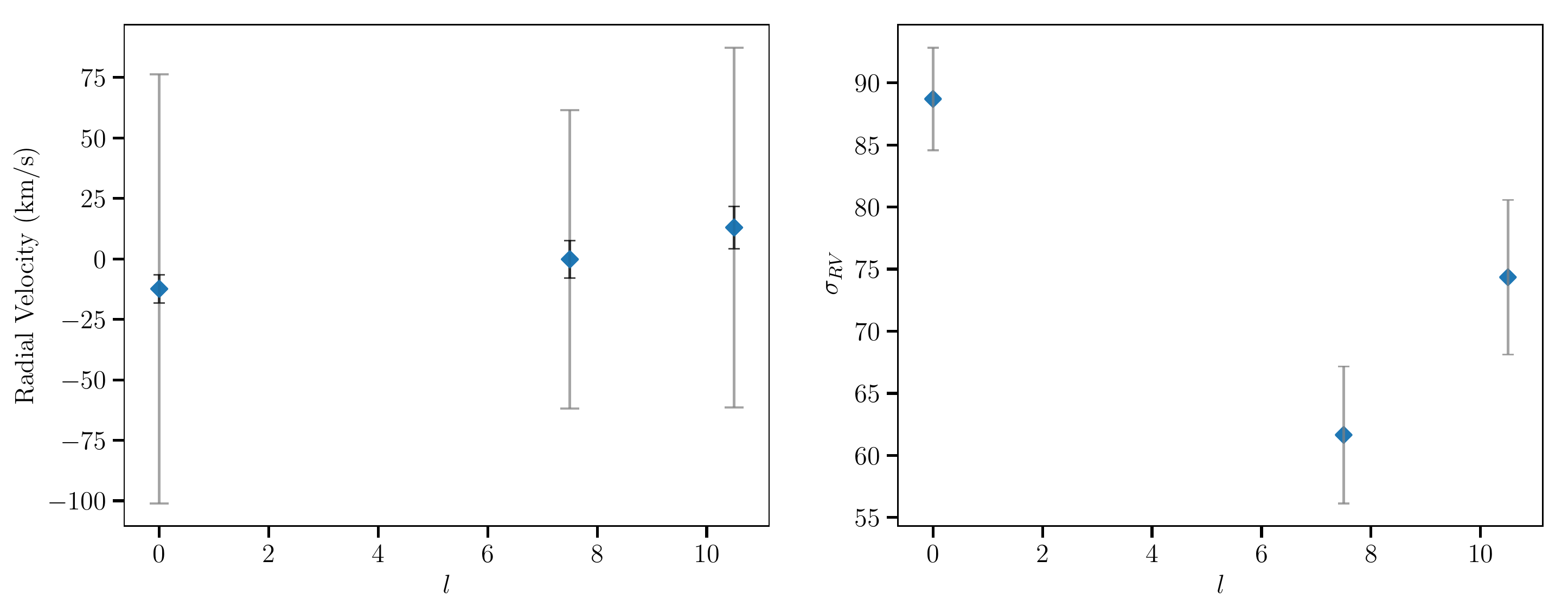}
    \caption{Left: the mean radial velocity as a function of longitude along the major axis at $b$ = 0$^{\circ}$. The light error bars show the 1$\sigma$ standard deviation around the mean age and the (smaller) dark error bars show the confidence on the mean measurement. Right: the 1$\sigma$ standard deviation of the mean in each field with error bars indicating the confidence on the measurement. There are 336 stars that comprise this figure.}
    \label{fig: RV along b=0}
\end{figure*}

Now we examine the stars within \rgal\ $\leq$ 3.5 kpc,  along the major axis, $b=0^\circ$, as shown in blue in Figure \ref{bulge map}. This includes 357 stars. We divide these stars into three fields, of $-3^\circ<b<3^\circ$ and $-7^\circ<l<7^\circ$, $7^\circ<l<10^\circ$, and $10^\circ<l<13^\circ$. We refer to these fields as the $(l, b)$ = $(0^{\circ}, 0^{\circ})$, $(7.5^{\circ}, 0^{\circ})$, and $(10.5^{\circ}, 0^{\circ})$ fields, respectively.
\par
The left-hand panel of Figure \ref{fig: [Fe/H] along b=0} shows the mean metallicity along the major axis, as calculated for each of the three fields described above. At $b=0^\circ$, there is a marginal increase in metallicity as $l$ increases. The right-hand panel of Figure \ref{fig: [Fe/H] along b=0} shows the 1-$\sigma$ standard deviation of the [Fe/H]. We find that the standard deviation markedly decreases as $l$ increases.
\par
Figure \ref{fig: ages along b=0} shows the mean age and age dispersion along the major axis. We do not observe any significant gradient in the mean age along the major axis $b=0$, as shown in the left-hand panel of the figure. The right-hand panel of Figure \ref{fig: ages along b=0} shows the 1$\sigma$ standard deviation of the age in the three fields. This is lowest for the $(l, b) = (0^{\circ}, 0^{\circ})$ field and fairly similar for the other two fields, which are about 15\% higher. 
\par
Figure \ref{fig: RV along b=0} shows the mean radial velocity and velocity dispersion along the major axis. We observe an increase in mean radial velocity as $l$ increases, similar to the one found in the metallicity in Figure \ref{fig: [Fe/H] along b=0}. We also notice that the $(l, b) = (0^{\circ}, 0^{\circ})$ field has a notably higher velocity dispersion than the other two fields, and the $(l, b) = (7.5^{\circ}, 0^{\circ})$ field has a lower velocity dispersion than the $(l, b) = (10.5^{\circ}, 0^{\circ})$ field. This is inversely correlated with the age dispersion, where the $(l, b) = (0^{\circ}, 0^{\circ})$ field has a much lower age dispersion than the other two fields, and the $(l, b) = (7.5^{\circ}, 0^{\circ})$ field has a higher age dispersion than the $(l, b) = (10.5^{\circ},0^{\circ})$ field. 

\subsection{Age Distribution across the Bulge}
\label{sec: bulge}

\begin{figure}[!ht]
    \centering
    \includegraphics[scale=0.55, keepaspectratio]{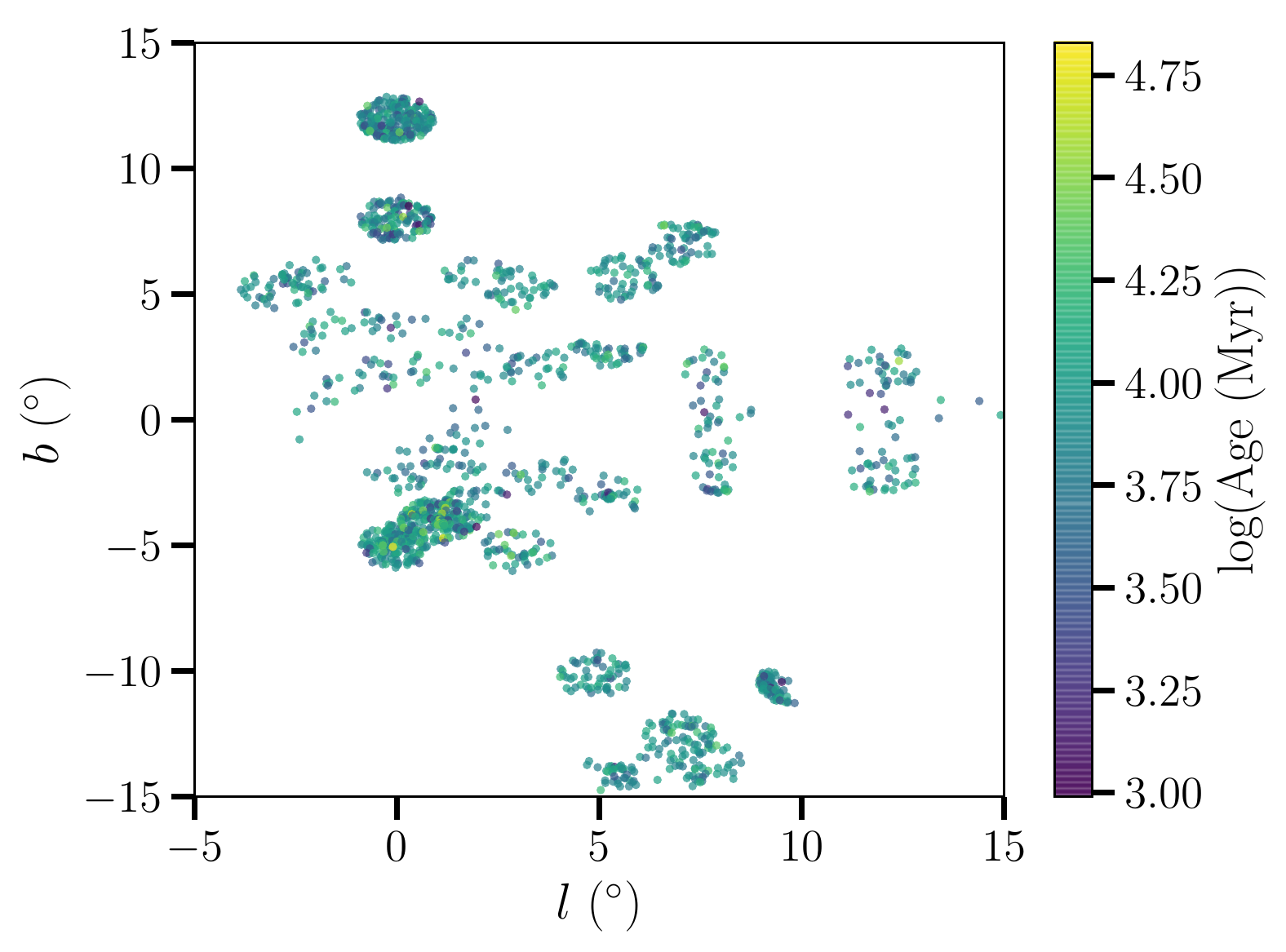}
    \caption{Age map of stars in the bulge region, defined as $(-15^\circ, -15^\circ)<(l, b)<(15^\circ, 15^\circ)$ and within \rgal\ $<$ 3.5 kpc of the Galactic center. There are 1654 stars in 39 unique fields in the region depicted.}
    \label{fig: age map}
\end{figure}

We now look at all the stars in the bulge region, which we define as stars within 3.5 kpc of the Galactic center and with $(-15^\circ, -15^\circ)<(l, b)<(15^\circ, 15^\circ)$. This yields a total of 1654 stars across 39 unique \apogee\ fields. In Figure \ref{fig: age map}, we plot these stars over the bulge region and color each by its age. We do not see any dramatic age gradients across the entire bulge region.
\par
\begin{figure}[!t]
    \centering
    \includegraphics[width=0.47\textwidth, keepaspectratio]{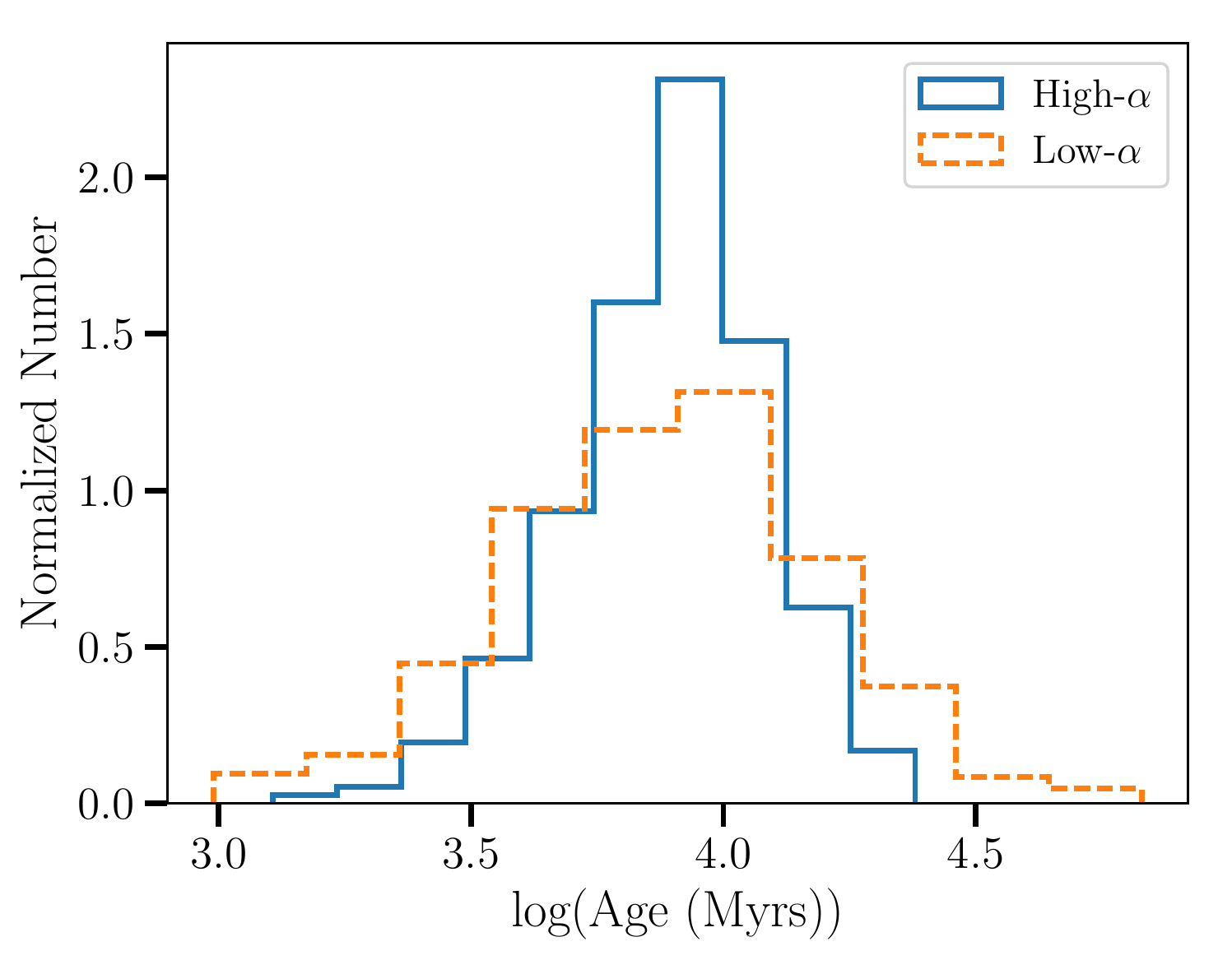}
    \caption{The age distribution of high-$\alpha$ and low-$\alpha$ stars in the bulge region, i.e. with $(-15^\circ, -15^\circ)<(l, b)<(15^\circ, 15^\circ)$ and \rgal\ $<$ 3.5 kpc. Of a total of 1654 bulge stars, 1203 are high-$\alpha$ and 451 are low-$\alpha$. The high-$\alpha$ stars have a distribution with mean $\log\tau$ = 3.89 dex and standard deviation $\sigma_{\log\tau}$ = 0.20 dex. The low-$\alpha$ stars have a distribution with mean $\log\tau$ = 3.88 dex and standard deviation $\sigma_{\log\tau}$ = 0.31 dex.}
    \label{fig: high vs low alpha age hist}
\end{figure}

For using \textit{The Cannon}, we further divide the stars into the high- and low-$\alpha$ populations that we defined previously. This yields 1203 high-$\alpha$ and 451 low-$\alpha$ stars in the bulge region. We plot the age distributions of the high- and low-$\alpha$ bulge stars in Figure \ref{fig: high vs low alpha age hist}. The two populations show similar mean ages ($\tau \approx$ 8 Gyr), but the low-$\alpha$ stars have a significantly wider ($\approx$40\%) age distribution.

\subsection{Age Distribution in the Bulge Compared to the Disk}
\label{sec: compare to disk}

Finally, we compare the age distribution of the bulge region to the disk. For this purpose, we define the disk stars as those between 7-9 kpc from the Galactic center and within 1 kpc of the Galactic plane, the solar neighborhood. For the bulge, we start with the same bulge stars as described in Section \ref{sec: bulge} and select only those within 1 kpc of the Galactic plane so that we have a consistent cut in $|z|$. This yields 35,343 disk stars and 1129 bulge stars for our purposes. 
\par
In the left panel of Figure \ref{fig: bulge vs disk ages}, we plot the age distribution of the bulge and the disk. In the right panel of the same figure, we plot the metallicity distribution of the bulge and the disk. We observe that the bulge has a broader metallicity range but a narrower age distribution than the solar neighborhood, both by about 30\%. The bulge stars have a higher mean age than the disk stars, at about 8 Gyr compared to 4 Gyr. 
\par
\begin{figure}[!t]
    \centering
    \includegraphics[width=0.47\textwidth, keepaspectratio]{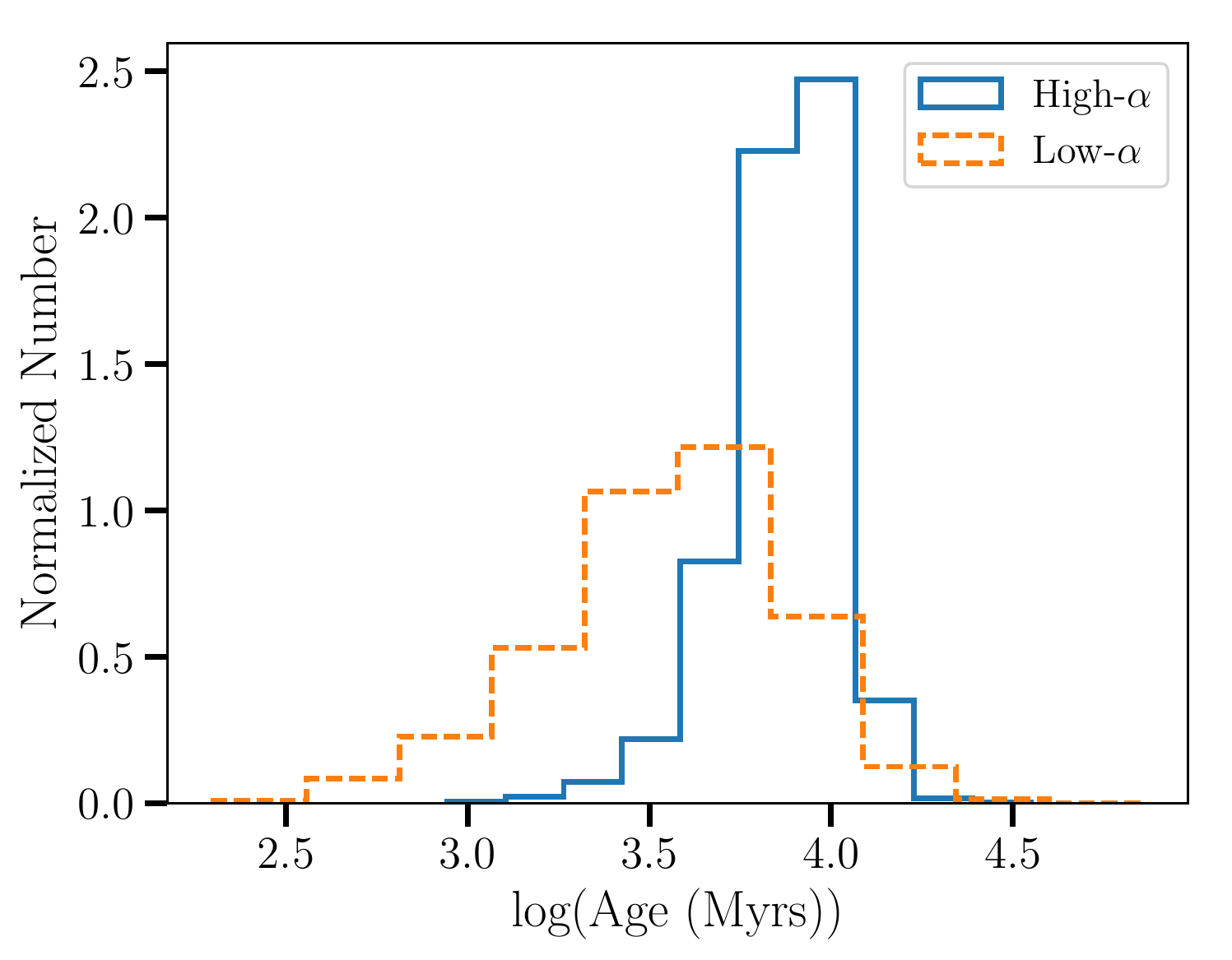}
    \caption{The age distribution of high-$\alpha$ and low-$\alpha$ stars in the disk near the solar neighborhood, i.e. with 7 kpc $<$ \rgal\ $<$ 9 kpc and $|z|<1$. Of a total of 35,343 bulge stars in this region, 6348 are high-$\alpha$ and 28,995 are low-$\alpha$. The high-$\alpha$ stars have a distribution with mean $\log\tau$ = 3.87 dex and standard deviation $\sigma_{\log\tau}$ = 0.16 dex. The low-$\alpha$ stars have a distribution with mean $\log\tau$ = 3.56 dex and standard deviation $\sigma_{\log\tau}$ = 0.33 dex.}
    \label{fig: high vs low alpha age hist for near sun}
\end{figure}

We found that when we divided stars into the high- and low-$\alpha$ populations in the inner Galaxy (\rgal\ $<$ 3.5 kpc), the two populations showed similar mean ages ($\tau \approx$ 8 Gyr). In Figure \ref{fig: high vs low alpha age hist for near sun}, we show the age distributions for the high and low-$\alpha$ populations near the Sun. In the solar neighborhood, the mean age of the low-$\alpha$ stars is $\tau=$ 3.5 Gyr, which is about half of that of the high-$\alpha$ stars near the Sun, at $\tau=$ 7.5 Gyr. So, not only is there a significant gradient in stellar age as a function of radius, with age decreasing from the inner to outer Galaxy, but this gradient is significantly higher, and primarily coming from the low-$\alpha$ population, which is old in the inner region.

\section{Discussion and Conclusions}

We have used the data-driven approach of \textit{The Cannon} to obtain ages and abundances for 125,367 \apogee\ stars. We provide the catalog of these here and undertake an analysis of the 2567 stars within the bulge region of \rgal\ $<$ 3.5kpc. For this analysis, we have examined the age, metallicity, and  abundance distributions, in particular focusing on differences in these as a function of longitude and latitude along the minor and major axes. We investigated the red clump stars that have been targeted in two high-latitude fields, to test for signatures of the X-shape. We have also contrasted the age and metallicity distributions for stars in the bulge, compared to the disk.  The main conclusions we draw from our analyses are as follows.

\begin{figure*}[!ht]
    \centering
    \includegraphics[width=0.9\textwidth, keepaspectratio]{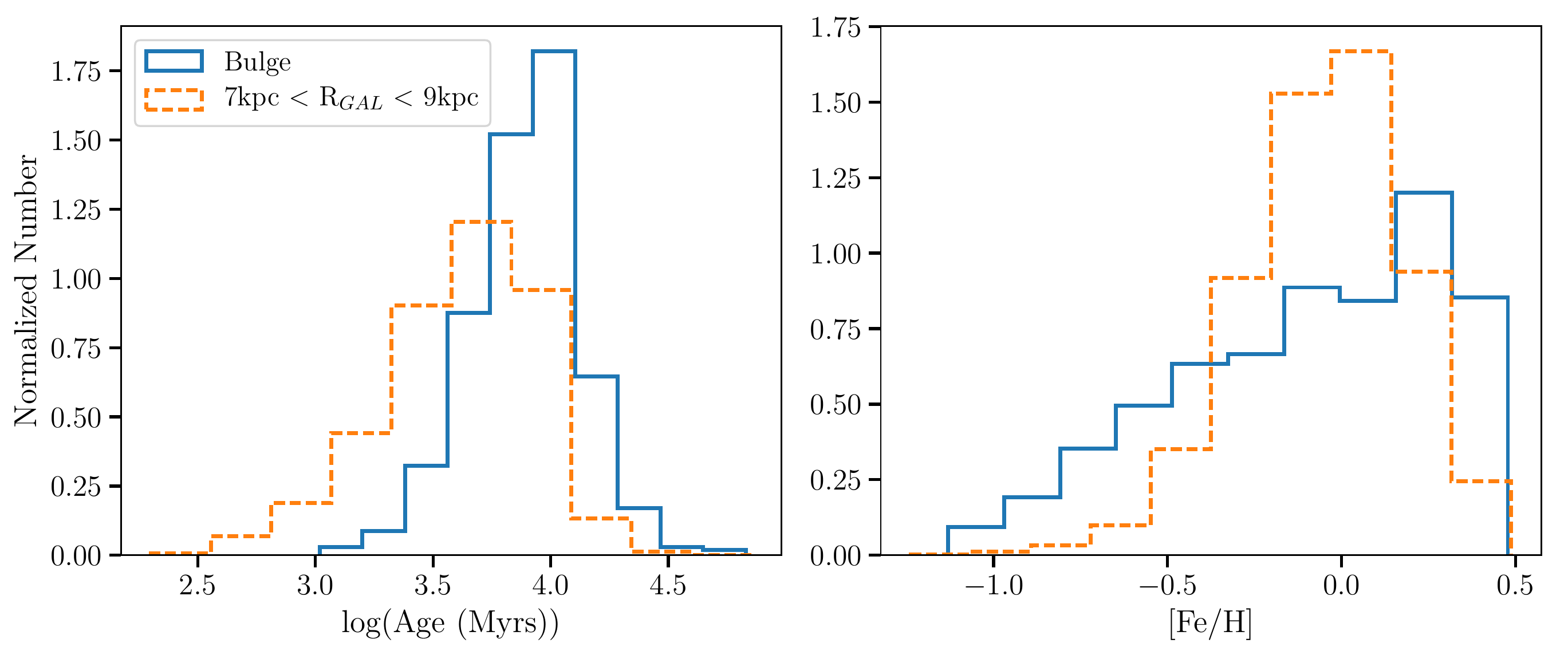}
    \caption{Right: distributions of age derived by \textit{The Cannon} for bulge stars, defined as stars with $(-15^\circ, -15^\circ)<(l, b)<(15^\circ, 15^\circ)$, \rgal\ $<$ 3.5 kpc, and $|z|<1$ ($N=1129$), and stars in the solar neighborhood, with 7 kpc $<$ \rgal\ $<$ 9 kpc and $|z|<1$ ($N=35343$). At left, the same for [Fe/H] derived by \textit{The Cannon}. The bulge has an age distribution with mean $\log\tau$ = 3.90 dex and standard deviation $\sigma_{\log\tau}$ = 0.24 dex, while the region near the sun has an age distribution with mean $\log\tau$ = 3.61 dex and standard deviation $\sigma_{\log\tau}$ = 0.33 dex. For [Fe/H], the bulge has a distribution with mean [Fe/H] = -0.10 dex and standard deviation $\sigma_{\textrm{[Fe/H]}}$ =  0.38 dex while the region near the sun has mean [Fe/H] = -0.05 dex and standard deviation $\sigma_{\textrm{[Fe/H]}}$ = 0.23 dex.}
    \label{fig: bulge vs disk ages}
\end{figure*}

\begin{enumerate}[noitemsep,parsep=0pt,partopsep=0pt,leftmargin=*]
    \item We see that the bulge is significantly older than the disk, with a mean age of 8 Gyr compared to 4 Gyr (calculated for stars within $|z|<1$ kpc). Many photometric studies of the bulge have demonstrated that the bulge is primarily old \citep[e.g.,][]{Ortoani1995, Zoccali2003, Clarkson2011, Renzini2018}. Spectroscopic studies have, however, revealed a fraction of young stars to be present \citep[i.e.,][]{Bensby2010, Bensby2013, Bensby2017}. Our results, of an age distribution of primarily old stars, but with a tail including younger stars, is broadly consistent with these prior findings. We have an age error from \textit{The Cannon} of about 0.2 dex for both the high- and low-$\alpha$ subsets. In Figure \ref{fig: bulge vs disk ages}, we see that the bulge has an age distribution with $\sigma_{\log\tau}$ = 0.24 and the disk has an age distribution with $\sigma_{\log\tau}$ = 0.33. We can calculate a more precise upper limit for the underlying age dispersion (but convolved with the age measurement error of the reference objects which is approximately propagated from training to test) of each region using the cross-validation age error from \textit{The Cannon}: \centerline{$\sigma_{underlying} = \left(\sigma_{distribution}^2 - \sigma_{Cannon}^2\right)^{1/2} $} This yields an underlying age dispersion of $\sigma_{\log\tau}$ =  0.13 dex in the bulge and $\sigma_{\log\tau}$ 0.26 dex in the disk, which corresponds to an age dispersion of $\sigma_{\log\tau}$ $\approx$3 Gyr for the disk and bulge, respectively. 
    \item Considering all stars in our fields within \rgal\ $<$ 3.5kpc, we find a negative age gradient with \rgal\ at low latitudes and a positive age gradient with \rgal\ at high latitudes. This may be a consequence of examining different populations;  the inside-out forming thin disk and the older, vertically extended, thicker disk population \citep[e.g.][]{ Debattista2017, Clarke2019,Ted2019b}.
    \item From the abundances for bulge stars within \rgal\ $<$ 3.5kpc, we see that, similarly to the solar neighborhood, the low-$\alpha$ stars are found preferentially near the plane and the high-$\alpha$ stars at larger heights from the plane. There is an age gradient that we see in the bulge with age decreasing from the oldest stars in the inner region to younger stars in the outer region. This is a consequence of inside-out formation  \citep[e.g. see][]{Frankel2019}. We see that looking at the low-$\alpha$ and high-$\alpha$ stars separately, only the low-$\alpha$ stars decrease in their mean age from bulge to solar neighborhood. The high and low-$\alpha$ stars in the bulge have similar mean ages (around 8 Gyr). However, the high and low-$\alpha$ stars around the Sun have mean ages of 8 Gyr and 4 Gyr, respectively. The different ages that we see across the [Fe/H]-[$\alpha$/Fe] plane, in the inner compared to the outer region, places strong constraints on models of the disk's formation.
    \item Two fields on the minor axis, at $(l, b) = (0^\circ, 8^\circ)$ and $(0^\circ, 12^\circ)$, were specifically targeted for red clump stars, which were further isolated using a $\log{g}$ cut using the inferred stellar parameters. These red clump stars show a gap in $K$-band apparent magnitude in the $(l, b) = (0^\circ, 8^\circ)$ field, but not at $(l, b) = (0^\circ, 12^\circ)$. This gap indicates that the X-shaped structure in the bulge is no longer present along the line of sight at the higher latitude, $(l, b) = (0^\circ, 12^\circ)$.
    \item Along the minor axis, the mean metallicity decreases fairly smoothly with galactic latitude, $b$. The mean ages show a less clear trend, and the ages are higher at negative latitudes compared to positive ones. The age and metallicity dispersion are scattered across latitude along the minor axis, with no clear correlation between the level of scatter in age and that in metallicity. However, the field with the lowest dispersion in both age and metallicity is that at the highest latitude, outside of the X-shape structure, at $(l, b) = (0^\circ, 12^\circ)$.
    \item Along the major axis, the dispersion in metallicity decreases significantly moving away from the Galactic center, as does the line-of-sight velocity dispersion. Conversely, the age dispersion is significantly smaller in the Galactic center compared to larger longitudes.
\end{enumerate}

In summary, the bulge is an old population with an age of $\tau$ $\approx$ 8 $\pm$ 3 Gyr, with a large range in metallicity. The Galactic center shows the largest spread in metallicity and smallest spread in age compared to farther out in $l$ and $b$ and radius, in the disk. The signature of inside-out formation in the disk is present in the low-$\alpha$ stars, but not the high-$\alpha$ ones. Our ensemble of results constrain models of Galactic formation and link directly to simulations, via our measure of ages.
\par
The number of stars in the bulge region will significantly increase with the public data release of \apogee\ DR16 and \sloanv's \mwm\ survey. The approach we have implemented in order to derive abundances and ages as well as our analysis is directly transferable to these new data, which will include many hundreds of thousands of stars in the bulge region.

\section{Acknowledgements} 

This research was funded by a Summer Undergraduate Research Fellowship from the Student-Faculty Programs office at Caltech. Melissa Ness is in part supported by a Sloan Foundation Fellowship. We thank the anonymous referee for their constructive feedback which improved the paper and Marina Rejkuba for helpful comments.

\bibliography{mknbib}

\end{document}